\documentclass[11pt,a4paper]{article}
\usepackage{jcappub}
\usepackage{eurosym}
\usepackage{graphicx}
\usepackage{float}
\usepackage{amssymb}
\usepackage{amsmath}
\usepackage{amsfonts}
\usepackage{amsbsy}
\usepackage{color}
\usepackage{rotating}
\usepackage[normalem]{ulem}

\renewcommand{\[}{\begin{equation}}
\renewcommand{\]}{\end{equation}}

\begin{document}

\title{Hidden in the background: A local approach to CMB anomalies}

\author{Juan C. Bueno S\'anchez}
\emailAdd{juan.c.bueno@correounivalle.edu.co}
\affiliation{Centro de Investigaciones en Ciencias  B\'asicas y Aplicadas, Universidad Antonio Nari\~no, Cra 3 Este \# 47A-15, Bogot\'a D.C. 110231, Colombia.}
\affiliation{Departamento de F\'isica, Universidad del Valle, A.A. 25360, Santiago de Cali, Colombia.}
\affiliation{Escuela de F\'isica, Universidad Industrial de Santander, Ciudad Universitaria, Bucaramanga 680002, Colombia.}

\abstract{We investigate a framework aiming to provide a common origin for the large-angle anomalies detected in the Cosmic Microwave Background (CMB), which are hypothesized as the result of the statistical inhomogeneity developed by different isocurvature fields of mass $m\sim H$ present during inflation. The inhomogeneity arises as the combined effect of $(i)$ the initial conditions for isocurvature fields (obtained after a fast-roll stage finishing many $e$-foldings before cosmological scales exit the horizon), $(ii)$ their inflationary fluctuations and $(iii)$ their coupling to other degrees of freedom. Our case of interest is when these fields (interpreted as the precursors of large-angle anomalies) leave an observable imprint only in isolated patches of the Universe. When the latter intersect the last scattering surface, such imprints arise in the CMB. Nevertheless, due to their statistically inhomogeneous nature, these imprints are difficult to detect, for they become hidden in the background similarly to the Cold Spot. We then compute the probability that a single isocurvature field becomes inhomogeneous at the end of inflation and find that, if the appropriate conditions are given (which depend exclusively on the preexisting fast-roll stage), this probability is at the percent level. Finally, we discuss several mechanisms (including the curvaton and the inhomogeneous reheating) to investigate whether an initial statistically inhomogeneous isocurvature field fluctuation might give rise to some of the observed anomalies. In particular, we focus on the Cold Spot, the power deficit at low multipoles and the breaking of statistical isotropy.} 




\maketitle
\vspace{1cm}

\section{Introduction}
Thanks to a wealth of high precision cosmological observations, specially those obtained by the WMAP \cite{Peiris:2003ff,Bennett:2003bz,Spergel:2003cb,Spergel:2006hy,Bennett:2010jb,Komatsu10,Bennett12,Hinshaw:2012aka} and Planck missions \cite{Ade:2013nlj,Ade:2013ydc,Ade:2015xua,Ade:2015lrj,Ade:2015ava}, cosmological inflation is widely recognized as the simplest paradigm to generate the observed adiabatic, nearly scale-invariant, Gaussian spectrum of superhorizon fluctuations imprinted in the Cosmic Microwave Background (CMB). In particular, single-field models are clearly favored by data. Despite this great success, cosmological inflation still faces a number of difficulties, the most obvious one being the large class of models consistent with data, but with different implications for particle physics. Another less pressing difficulty is the persistence, for more than a decade now, of large-angle anomalies in the CMB, which suggests that single-field inflation might need an extension of some kind. These anomalies, currently accepted as real features of the data, were observed for the first time by the WMAP satellite \cite{Bennett:2003bz,Bennett:2010jb} and later confirmed by Planck \cite{Ade:2013nlj,Ade:2015hxq}. Since their existence seems to pose a relative challenge for single-field inflation, an important theoretical effort has been dedicated over the past decade to elucidate their origin (see \cite{Schwarz:2015cma} for a recent review). 

Since observations clearly support an adiabatic, nearly scale-invariant, Gaussian spectrum of superhorizon perturbations (according to the generic predictions of single-field inflation), here we take the view that the primordial perturbation spectrum is not only sourced by the inflaton, but also receives the contribution from other fields in the theory. This is the case, for example, of mixed inflaton-curvaton perturbations \cite{Easson:2010uw,Kinney:2012ik} or inhomogeneous reheating \cite{Dvali:2003em,Dvali:2003ar}. Moreover, since the existence of large-angle anomalies imply the breaking of the statistically homogeneity/isotropy of the CMB, and also since some of them can have a different origin (see for example \cite{Sarkar:2010yj}), in this paper we envisage them as the result of the statistical inhomogeneity obtained by different isocurvature fields during the last stage of slow-roll inflation. Our framework then hypothesizes with the existence of isocurvature field perturbations as the precursors of CMB anomalies, and that the latter are realized through different mechanisms using different isocurvature fields. Specifically, we make use of the curvaton mechanism (both scalar \cite{Lyth:2001nq,Lyth:2002my} and vector \cite{Dimopoulos:2006ms,Dimopoulos:2011ws}) and the inhomogeneous reheating to account for some more of the most robust anomalies appearing in the CMB sky: the Cold Spot, the power deficit at low $\ell$ and the breaking of statistical isotropy. Of course, depending on the specifics of the mechanism under consideration, the isocurvature perturbation may be either totally converted into a curvature perturbation, or partially converted, thus generating a residual isocurvature perturbation. 

In our setting, the development of the statistical inhomogeneity in the additional isocurvature fields owes to the combined effect of $(i)$ the initial condition for isocurvature fields at the onset of slow-roll inflation, $(ii)$ their inflationary fluctuations during slow-roll and $(iii)$ their interaction with other degrees of freedom present in the theory. Similar ideas, but leading to statistically homogeneous perturbations, have been explored in the literature using the inflaton instead of an isocurvature field. Well-known examples of this are based on the existence of a particle production mechanism, originating from the coupling of the inflaton to other fields in the theory, that modifies the perturbation spectrum of the inflaton \cite{Chung:1999ve,Green:2009ds,Langlois:2009jp,Barnaby:2009mc}. However, after triggering the production mechanism, the inflaton continues its rolling and returns to its slow-roll attractor. In contrast, the scenario considered in this paper is different in two aspects. In the first place, the particle production mechanism is triggered by an isocurvature field, and hence, the perturbation spectrum of the inflaton does not become modified. And secondly, once the production mechanism is triggered, the isocurvature field never recovers its previous dynamics, but becomes trapped similarly to a moduli field \cite{Kofman:2004yc}.

A most important aspect of the framework here discussed is the generation of the initial condition for isocurvature fields. As explained later on, in order for statistical inhomogeneity to arise, it is first necessary to assume a large field value at the onset of the slow-roll. The difficulty to motivate such a large value for scalar fields with mass $m\sim H$ is that they are expected to be of order $H$ \cite{Starobinsky:1994bd}, although this result applies when the scalar field is in its equilibrium state in de Sitter space. Despite this drawback, it was shown in \cite{Sanchez:2014cya} that to generate an initial condition appropriate for the development of large inhomogeneities in $\sigma$, it suffices to consider a sustained stage of non-slow-roll, or fast-roll inflation\footnote{Note that large isocurvature fluctuations can also arise in the slow-roll regime during $N$-flation \cite{Easther:2013rva,Price:2014xpa}.} \cite{Linde:2001ae}. However, fast-roll inflation cannot be reconciled with observations, for the curvature of the potential results in an excessive scale-dependence of the spectrum. Therefore, since observations clearly support slow-roll inflation as the origin of the primordial spectrum imprinted in the CMB, one is naturally driven to conclude that no significant departure from slow-roll becomes relevant to describe the primordial spectrum. Nevertheless, it is feasible that such departures leave an observable imprint, generating a power deficit in the low multipoles  \cite{Linde:2001ae,Contaldi:2003zv,Schwarz:2009sj,Boyanovsky:2009xh,Ramirez:2011kk,Lello:2013mfa} or oscillatory features in the power spectrum \cite{Achucarro:2010da,Achucarro:2012sm}. In any case, to maintain the agreement with observations, such departures must be sufficiently moderate.

Larger departures from slow-roll inflation, however, are required to produce an appropriate initial condition for isocurvature fields \cite{Sanchez:2014cya}. In turn, such departures are expected during the early stages of inflation, when the scalar potential is dominated by large K\"ahler corrections \cite{Copeland:1994vg,Dine:1995uk}. Indeed, on general grounds one can expect that inflation begins somewhat close to the Planck scale in some regime substantially away from slow-roll attractors \cite{Lyth:1998xn}. This may be the case, for example, if large supergravity corrections to the scalar potential do not cancel out with sufficient accuracy \cite{Copeland:1994vg,Dine:1995uk}. Owing to the curvature of the scalar potential, these departures must take place during \emph{primary} inflation \cite{Lyth:1998xn}, which is the epoch when the observable Universe is still inside the horizon. Since the perturbation spectrum cannot be probed on those scales, primary inflation is usually deemed as relatively uninteresting in comparison to the phase of (slow-roll) inflation during which cosmological scales exit the horizon. Nevertheless, here we challenge this attitude towards primary inflation and investigate the initial conditions that a stage of primary fast-roll inflation can generate and whether such initial conditions can leave an observable imprint in the primordial spectrum. In this sense, it is worth emphasizing that recent results provide a positive answer in this direction \cite{Chen:2013eaa,Wetterich:2015gya}, showing that if not too long-lasting, a primary phase of inflation may have consequences for the observed primordial spectrum. 

Another fundamental aspect of this research is the assessment of the probability that a single isocurvature field fluctuation becomes statistically inhomogeneous at the end of inflation. This probability, however, depends on the details of the primary phase and, consequently, a full computation requires a particular model of inflation. Although the discussion in this paper proceeds without specifying any particular model, we make an assumption (whose validity depends on the model of inflation) allowing us to carry out a computation of this probability. In any case, since the naturalness of our proposal suggests that this probability be sizable (as we find it to be for fields with $m\sim H$ under the appropriate circumstances), this model dependence offers an opportunity to use our framework as a tool to discriminate models of inflation. We defer a detailed search in this sense for future research.

The paper is organized as follows. In section 2 we study the evolution of a single isocurvature field during inflation, explaining the mechanism whereby the field becomes inhomogeneous at the end of inflation. In section 3 we elaborate on a modification to the stochastic approach to inflation aimed at studying the main features of the classical field distribution at the end of inflation. Moreover, we estimate the probability that the field becomes inhomogeneous at the end of inflation. In section 4 we apply the curvaton mechanism (scalar and vector) and the inhomogeneous reheating in order to account for some of the CMB anomalies. We present our conclusions in section 5.

\section{Inflationary growth of spectator fields}\label{sec13}
We describe now the evolution of a general isocurvature field from the beginning of inflation until then end of it. A most important stage during the evolution is the fast-roll, for it is during this phase that the field obtains a value significantly larger than the Hubble scale $H$ that is crucial for our framework. Although producing a classical condensate with a large value for a field of mass $m\sim H$ is an interesting prospect, we have to recall that in our scenario this production takes place during a primary phase, and hence the possibility exists that the classical field becomes negligible when the observable Universe exits the horizon. Therefore, to describe the evolution of the condensate during inflation we must specify the entire inflationary stage. To do so, we write the total length of inflation as
\[\label{eq7}
N_{\rm tot}=N_{\rm fr}+N_{\rm sr}\,,
\]
where the subscripts ``fr'' and ``sr'' stand for fast-roll and slow-roll, respectively. Here, we allow the primary phase of inflation to contain a slow-roll stage. In that case, slow-roll inflation lasts longer than demanded by observations and we write
\[\label{eq37}
N_{\rm sr}=N_{\rm sr}^p+N_*\,,
\]	
where $N_{\rm sr}^p>0$ denotes the length of the primary slow-roll phase and $N_*$ is the number of $e$-foldings demanded by observations, typically in the interval $40\lesssim N_*\lesssim 70$. We emphasize that the situation here examined, i.e. a non-negligible primary phase of slow-roll between the fast-roll stage and the time of horizon crossing, is in clear contrast to the one usually considered in models hypothesizing the existence of fast-roll stage to account, for example, to the power deficit at low $\ell$ \cite{Linde:2001ae,Contaldi:2003zv,Schwarz:2009sj,Boyanovsky:2009xh,Ramirez:2011kk,Lello:2013mfa}. 

Since the perturbation spectrum cannot be probed on scales that exited the horizon during primary inflation, the latter is mostly unconstrained by observations. Then, in principle one might consider an arbitrary length $N_{\rm sr}^p$. A minimal requirement on $N_{\rm sr}^p$, however, stems from the fact that we focus on a fast-roll stage that constitutes a large departure from slow-roll. Given the excellent agreement between CMB observations and the slow-roll paradigm, we must afford at least a few $e$-foldings between the end of the fast-roll stage and the time of horizon crossing for cosmological scales. We take this transition to be included in the first $e$-foldings of primary slow-roll, and hence $N_{\rm sr}^p>{\cal O}(1)$. Apart from this, there is no upper bound on $N_{\rm sr}^p$, which might be set arbitrarily large. Of course, the archetypical example in this case is eternal inflation \cite{Linde:1986fd}. Nevertheless, in this paper we will restrict ourselves to the case when the primary slow-roll phase is relatively short-lived, with $N_{\rm sr}={\cal O}(10^2)$. As suggested before, the main reason for this owes to our intent to focus on fluctuating fields with masses $m\sim H$. Indeed, if the primary phase of slow-roll is too long-lasting, the natural expectation is that when the observable Universe exits the horizon the fluctuations of all such fields will have their equilibrium amplitude in de Sitter space \cite{Starobinsky:1994bd}, thus erasing the memory of the initial condition generated during the fast-roll stage.

\subsection{A sustained stage of fast-roll inflation}\label{sec1}
According to our previous discussion, we consider a phase with a non-negligible variation of the Hubble parameter according to $\dot H=-\epsilon H^2$, where $\epsilon>0$ is kept constant for simplicity\footnote{Note that a strictly constant $\epsilon$ cannot be consistently obtained in single-field slow-roll inflation, and hence some sort of multifield dynamics is impicitly assumed here. A particular realization of this dynamics has been recently discussed in the context of higher-dimensional inflation, where an attractor solution with a constant $\epsilon\geq1/2$ is reported \cite{Burgess:2016ygs}.}. Although the analysis below is valid for any $\epsilon$, we are mainly interested in the case when $\epsilon$ is relatively large, but still consistent with inflation. Keeping $\epsilon$ constant, it is straightforward to obtain the background evolution
\[\label{eq18}
H=H_0a(t)^{-\epsilon}\,\,,\,\, a(t)=(1+\epsilon H_0t)^{1/\epsilon}\,,
\]
where $H_0$ is the Hubble parameter at the beginning of inflation. In this background, we consider a massive, free scalar field $\sigma$ minimally coupled to gravity with Lagrangian
\[
{\cal L}=\frac12\partial_\mu\sigma\partial^\mu\sigma-V(\sigma)\,,
\]
where $V(\sigma)$ is the effective scalar potential. In our setting, $\sigma$ is a generic isocurvature field present during inflation, and hence its energy density does not affect the inflationary background. In the following we write $V(\sigma)=\frac12m_\sigma^2\sigma^2$, with $m_\sigma^2=c_\sigma H^2$, and pay special attention to the case $c_\sigma={\cal O}(10^{-1})$, thus implying that $m_\sigma$ is dominated by the Hubble-induced correction. 

The evolution equation for the perturbation modes of $\sigma$ is
\[\label{eq1}
\delta\ddot\sigma_k+3H\dot\delta\sigma_k+\left(\frac{k^2}{a^2}+c_\sigma H^2\right)
\delta\sigma_k=0\,.
\]
Imposing the Bunch-Davies vacuum in the subhorizon limit $k/aH\to0$, the solution to Eq.~(\ref{eq1}) is
\[\label{eq5}
\delta\sigma_k(t)=a^{-1/2}e^{i\pi(\nu+1/2)/2}\sqrt{-\frac{\pi\tau}4}
\,H_\nu^{(1)}\left(-k\tau\right)\,,
\]
where $\tau=-[(1-\epsilon)aH]^{-1}$ is the conformal time and
\[\label{eq40}
\nu^2\equiv\frac94-\frac{c_\sigma-\epsilon(3-2\epsilon)}{(1-\epsilon)^2}\,.
\]
In the superhorizon limit, we find $\delta\sigma_k\propto a^{3/2-\nu+\epsilon(\nu-1/2)}$. Expanding to first order in $c_\sigma$, we have $\delta\sigma_k\propto a^{-c_\sigma/(3-\epsilon)}$, and hence field perturbations evolve in the timescale \mbox{$\tau_\sigma\equiv\frac{3-\epsilon}{c_\sigma}\,H^{-1}$}.

Using Eq.~(\ref{eq5}) we obtain the perturbation spectrum
\[\label{eq57}
{\cal P}_{\delta\sigma}(k)\equiv\lim_{k/aH\to0}\frac{k^3|\delta\sigma_k|^2}{2\pi^2}=\gamma\,\frac{H^2}{4\pi^2}\left(\frac{k}{aH}\right)^{3-2\nu}\,,
\]
where $\gamma\equiv\frac{2^{-1+2\nu}\Gamma(\nu)^2}{\pi(1-\epsilon)^{1-2\nu}}$. Assuming now that $\sigma=0$ at the beginning of inflation, we obtain the field variance
\[\label{eq10}
\Sigma^2(N,c_\sigma,\epsilon)\equiv\langle(\sigma-\bar\sigma)^2\rangle=\gamma\frac{H^2}{4\pi^2(3-2\nu)}\left(1-e^{-(3-2\nu)N}\right)\,,
\]
where $N$ is the number of elapsed $e$-foldings from the beginning of inflation. Since $H=H(N)$, at the end of the fast-roll stage we have $H(N_ {\rm fr})=H_*$, where $H_*$ is the Hubble parameter during slow-roll inflation.

\begin{figure}[htbp]
\centering\epsfig{file=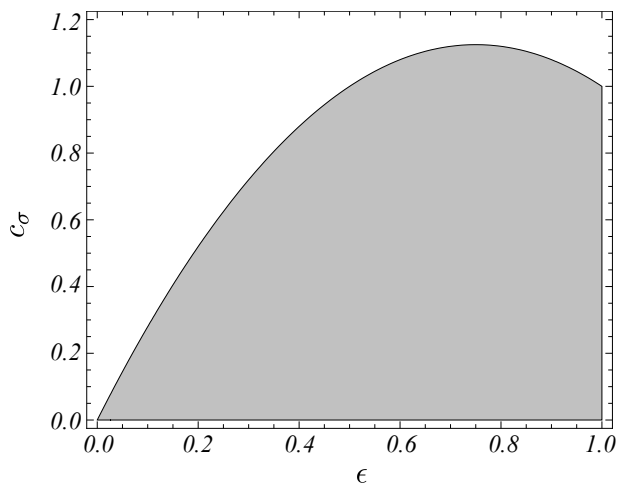,width=8cm}
\caption{Plot of the parametric region $3-2\nu<0$ leading to the unstable growth of $\Sigma^2$.}\label{fig6}
\end{figure}

The situation of interest to us is when $\epsilon$ is relatively large, but keeping $\epsilon<1$ to have inflationary expansion. In that case, $H$ does not remain approximately constant, but changes in the timescale $\tau_H\equiv\epsilon^{-1}H^{-1}$. If $c_\sigma$ is not too large we can have $c_\sigma<\epsilon$ so that $\tau_H<\tau_\sigma$, or $3-2\nu<0$ equivalently. In Fig.~\ref{fig6} we plot the parametric range corresponding to $3-2\nu<0$ (shaded region). The plot shows that even for relatively massive fields, up to $m_\sigma\simeq H$, an effective tachyonic instability develops. Although $m_\sigma^2>0$, the fact that $\tau_H<\tau_\sigma$ implies that field fluctuations evolve slower than $H$, and hence the ratio $\Sigma^2/H^2$ grows unbounded. As a result, at the onset of slow-roll inflation, when $H$ becomes approximately constant, the amplitude of the field fluctuations produced during the fast-roll can be so much larger than their corresponding equilibrium value in the slow-roll regime. Therefore, when $\epsilon$ is sufficiently large, field fluctuations go \textit{out-of-equilibrium}. Moreover, since a fluctuation produced during the fast-roll (whose magnitude is determined by $H$) becomes larger than those produced later, the amplitude of field fluctuations at the onset of the slow-roll is dominated by those produced at the beginning of the fast-roll stage, and hence is mostly determined by the Hubble scale at the beginning of inflation. 

\begin{figure}[htbp]
\centering\epsfig{file=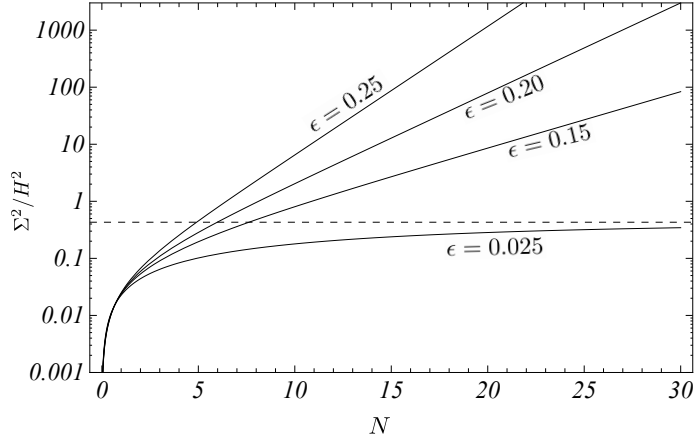,width=9cm}\caption{Evolution of $\Sigma^2/H^2$ during inflation for $c_\sigma=0.15$ and several values of $\epsilon$, as indicated. The dashed line represents the amplitude of the equilibrium fluctuations, as obtained from Eq.~(\ref{eq10}) for $\epsilon=0.025$. Only for this curve $3-2\nu>0$.}\label{fig3b}
\end{figure}

The growth of $\Sigma^2$ during inflation is illustrated in Fig.~\ref{fig3b}, where we take $c_\sigma=0.15$ and plot the behavior for different values of $\epsilon$. When $\epsilon$ is sufficiently small, the growth of $\Sigma^2/H^2$ becomes limited by an upper bound. This is exemplified for $\epsilon=0.025$. For larger $\epsilon$, corresponding to $2\nu>3$, our plot evidences the unstable growth of $\Sigma^2$. As a result, and contrary to the expectation in slow-roll inflation, it becomes perfectly possible to obtain classical values of $\sigma$ well above $H$ even for fields with $m_\sigma\sim H$. The price to pay, however, is the existence of a sustained stage of fast-roll inflation.

\subsection{Evolution during slow-roll}\label{sec6}
We review now the slow-roll evolution of the classical field $\sigma$ generated during the fast-roll stage. Since the classical field continues to fluctuate during the subsequent slow-roll phase, we must take into account the corresponding buildup of fluctuations on $\sigma$. We do so by resorting to the well-known stochastic approach to inflation \cite{Starobinsky:1986fx,Sasaki:1987gy,Starobinsky:1994bd}. As before, we consider $\sigma$ to be a non-interacting field and take $\dot H=0$ during slow-roll inflation. Any later appearance of $\epsilon$ will refer to the one characterizing the fast-roll stage.

The equation of motion for the homogeneous part of $\sigma$ is
\[
\ddot\sigma+3H\dot\sigma+\frac12\,c_\sigma H^2\sigma=0\,.
\]
With $\dot H=0$, the growing mode solution is
\[
\sigma\propto a(t)^{-3/2+\sqrt{9/4-c_\sigma}}\simeq a(t)^{-c_\sigma/3}
\]
to first order in $c_\sigma$. 
Denoting by $\sigma_{\rm sr}$ the field value at the onset of slow-roll inflation in the horizon-sized patch from which our observable Universe emerges, we have 
\[\label{eq72}
\sigma\simeq\sigma_{\rm sr}e^{-c_\sigma N/3}\,,
\]
where $N$ counts the number of $e$-foldings from the beginning of the slow-roll. Since we intend to focus on $c_\sigma={\cal O}(10^{-1})$, the motion of $\sigma$ is close to critically damped. Then, strictly speaking, the field cannot be said to be in slow-roll. Indeed, although $c_\sigma={\cal O}(10^{-1})$ implies little evolution of $\sigma$ while CMB scales are exiting the horizon, this is certainly not the case when we track the evolution of $\sigma$ until the end of inflation. In turn, it is precisely the latter that plays an important role in our framework.

To take into account the influence of inflationary fluctuations on the dynamics of the classical field we make use of the stochastic approach to inflation \cite{Starobinsky:1986fx,Sasaki:1987gy,Starobinsky:1994bd} from the onset of the slow-roll (at $t=t_{\rm sr}$) until the end of inflation (at $t=t_{\rm end}$). For a classical field of constant mass $m_\sigma<H$, the evolution of its associated probability density is described by the Fokker-Planck equation \cite{Starobinsky:1994bd}
\[\label{eq15}
\frac{\partial P}{\partial t}=\frac{\partial}{\partial \sigma}
\left(\frac{V'(\sigma)}{3H}\,P\right)+\frac12{\cal D} \frac{\partial^2P}{\partial \sigma^2}\,,
\]
where ${\cal D}=\frac{H^3}{4\pi^2}$ is the diffusion coefficient. To solve for it, we impose the initial condition
\[\label{eq38}
P(\sigma,t_{\rm sr})=\delta(\sigma-\sigma_{\rm sr})\,.
\]
Since typical field values at the end of the fast-roll are of order $\Sigma(N_{\rm fr})$, our previous results motivate us to consider $\sigma_{\rm sr}\gg H$. As for boundary conditions, the standard approach to stochastic inflation assumes that $P(\sigma,t)$ evolves in unbounded field space. In that case, the conservation of the probability density demands the boundary conditions
\[\label{eq24}
P(\pm\infty,t)=0,\quad\partial_\phi P(\pm\infty,t)=0\,.
\]
Then, the solution to Eq.~(\ref{eq15}) is well approximated by a Gaussian distribution with mean and variance given by
\[\label{eq36}
\bar\sigma(t)=\sigma_{\rm sr}e^{-c_\sigma N/3}\quad,\quad
\Sigma^2=\frac{3H^4}{8\pi^2m_\sigma^2}
\left[1-e^{-2c_\sigma N/3}\right]\,.
\]

\begin{figure}[htbp]
\centering\epsfig{file=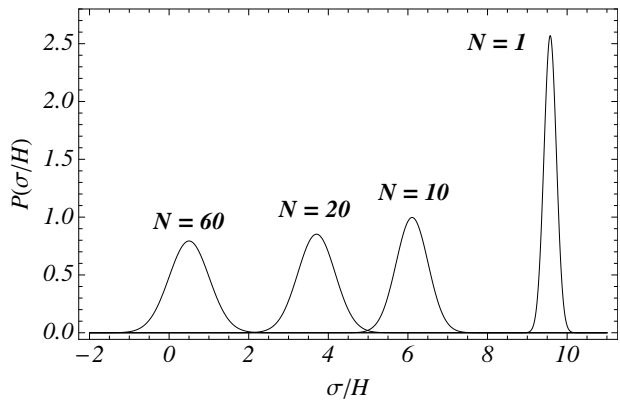,width=8cm}\caption{Snapshots of the probability density $P(\sigma,t)$ taken during the slow-roll stage $N$ $e$-foldings after the initial time (as indicated). We set the initial value to $\sigma_{\rm sr}=20\sigma_{\rm eq}$.}\label{fig1}
\end{figure}

We exemplify the evolution of $P(\sigma,t)$ in Fig.~\ref{fig1}, setting the initial $\sigma_{\rm sr}$ larger than the amplitude of equilibrium fluctuations in de Sitter space, namely $\sigma_{\rm sr}^2>\sigma_{\rm eq}^2\equiv\frac{3H^2}{8\pi^2m_\sigma^2}$. For the purpose of illustration we choose $\sigma_{\rm sr}=20\sigma_{\rm eq}$, which can be conveniently justified by a previous stage of fast-roll inflation with $c_\sigma=0.15$, $\epsilon=0.2$ and $N_{\rm fr}\simeq20$, for example. The essential point to stress here is that the probable values of $\sigma$ remain within the same order of magnitude even if the probability density has not reached its equilibrium state. This can be shown by using the definition of $\Sigma^2$ in Eq.~(\ref{eq10}) to write any probable field value as $\sigma=\bar\sigma+\alpha \Sigma=\Sigma(\bar\sigma/\Sigma+\alpha)$, where $|\alpha|\lesssim1$. Therefore, we find $\sigma\sim\bar\sigma$ for all probable field values when $\bar\sigma>\Sigma$, whereas $\sigma\sim\Sigma$ for all probable values in any other case. As shown below, this may change dramatically when $\sigma$ couples to other degrees of freedom.

\subsection{The role of interactions}\label{sec8}
We investigate a system of two interacting, massive scalar fields $\sigma$ and $\chi$ minimally coupled to gravity and whose energy density remains always subdominant. Using an interaction term of the form $g^2\sigma^2\chi^2$ and ignoring the interactions of $\sigma$ and $\chi$ with other fields, the Lagrangian of the system is
\[\label{eq27}
{\cal L}=\frac12\partial_\mu\sigma\partial^\mu\sigma-V(\sigma)+\frac12\partial_\mu\chi\partial^\mu\chi
-\frac12 m_{0\chi}^2\chi^2-\frac12\,g^2\sigma^2\chi^2,
\]
where $g$ is a coupling constant and $m_{0\chi}$ is the bare mass of $\chi$. This interaction term is ubiquitous in quantum field theory, and its consequences have been extensively studied in the theory of reheating and preheating \cite{Dolgov:1982th,Abbott:1982hn,Traschen:1990sw,Dolgov:1989us,Kofman:1994rk,Shtanov:1994ce,Kofman:1997yn,Felder:1998vq}. Moreover, this coupling results in a trapping mechanism whereby points of enhanced symmetry become a preferred location for string moduli \cite{Kofman:2004yc,Watson:2004aq,Enomoto:2013mla}. This trapping mechanism has been employed in inflation model building (trapped inflation) \cite{Kadota:2003tn,BuenoSanchez:2006eq,BuenoSanchez:2006ah,Green:2009ds}, also to generate non-Gaussianity of the inflaton's perturbation spectrum \cite{Barnaby:2009mc,Wu:2006xp,Lee:2011fj} and, more recently, to study the stochastic evolution of coupled flat directions \cite{Sanchez:2012tk}.

In the Hartree approximation, the dynamics of $\sigma$ and $\chi$ is determined by the equations
\[\label{eq65}
\ddot\sigma+3H\dot \sigma+(c_\sigma H^2+g^2\langle\chi^2\rangle)\sigma=0
\]
and
\[\label{eq81}
\ddot\chi_k+3H\dot\chi_k+\left(\frac{k^2}{a^2}+m_{0\chi}^2+g^2\sigma^2\right)\chi_k=0\,,
\]
where 
\[
\langle\chi^2\rangle=\frac1{(2\pi)^3}\int|\chi_k|^2d^3k\,.
\]
The effective masses of $\sigma$ and $\chi$ are $m_\sigma^2=c_\sigma H^2+g^2\langle\chi^2\rangle$ and $m_\chi^2=m_{0\chi}+g^2\sigma^2\simeq g^2\sigma^2$, where we neglect the bare mass for simplicity. 

As discussed in \cite{Sanchez:2012tk,Sanchez:2014cya}, the evolution of $\sigma$ strongly depends on the magnitude of its initial value $\sigma_{\rm sr}$ with repect to the crossover value $\sigma_c\equiv\sqrt{10}g^{-1}H_*$. If $\sigma_{\rm sr}<\sigma_c$, we have $m_\chi^2<H^2$ and the $\chi$ field undergoes particle production during inflation, which then blocks the growth of fluctuations in $\sigma$ \cite{Enqvist:2011pt,Kawasaki:2012bk}. On the contrary, if $\sigma_{\rm sr}> \sigma_c$ the $\chi$ field becomes heavy and does not get produced during inflation, but contributes to the effective potential of $\sigma$ through quantum corrections. Since we take $m_\sigma$ to be dominated by the Hubble-induced correction, in the following we neglect the quantum corrections coming from the $\chi$ field. Therefore, during the slow-roll phase the field scales as
\[\label{eq20}
\sigma\propto a^{-c_\sigma/3}
\]
for as long as $\sigma>\sigma_c$. When $\sigma\leq\sigma_c$, the $\chi$ field becomes produced during inflation and the effective mass of $\sigma$ increases. As a result, $\sigma$ evolves faster towards $\sigma=0$, thus allowing the production of $\chi$ to continue. The outcome of this self-sustained process is that $\sigma$ ends up oscillating about $\sigma=0$ soon after its interaction with $\chi$ becomes dynamically important\footnote{In essence, this process is no different from the trapping one described in \cite{Kofman:2004yc,Watson:2004aq,Enomoto:2013mla}. In our case, however, the production of the $\chi$ field does not influence the dynamics of the inflaton, nor does it backreact on its perturbations, at least at horizon crossing.}. The typical field value during the oscillatory phase scales as
\[\label{eq22}
\sigma\propto a^{-3/2}\,.
\]
Owing to the inflationary fluctuations of $\sigma$ and to the sharp cutoff for the development of a classical field $\chi$, the onset of the oscillatory phase (or the trapping of $\sigma$) does not occur everywhere at the same time. As a result, it becomes conceivable to find regions of the observable Universe where $\sigma$ remains oblivious to its interactions, and hence in slow-roll (or critically damped) until the end of inflation and scaling as in Eq.~(\ref{eq20}), whereas in others the field is already oscillating by the end of inflation, thus scaling as Eq.~(\ref{eq22}). In the latter case, the typical value of $\sigma$ can be estimated by
\[\label{eq80}
\sigma(\mbox{\boldmath$x$})\sim\sigma_c\,\exp[-3N_{\rm osc}(\mbox{\boldmath$x$})/2]\,,
\]
where we introduce the stochastic variable $N_{\rm osc}(\mbox{\boldmath$x$})$ representing the remaining number of $e$-foldings at the onset of the oscillatory regime at the location $\mbox{\boldmath$x$}$. 

Depending on the model parameters, it is possible to arrange that $\sigma$ remains in its slow-roll stage until the end of inflation only in sparse regions of the Universe. Therefore, in a large fraction of the observable Universe, where $\sigma$ is already oscillating at the end of inflation, $\sigma$ becomes exponentially suppressed according to Eq.~(\ref{eq80}), whereas $\sigma$ retains a relatively large value $\sigma\sim\sigma_c$ in sparse regions of the Universe\footnote{It is also possible to tune parameters so that $\sigma$ is oscillating in the entire observable Universe at the end of inflation. But since $\sigma$ becomes exponentially small compared to $\sigma_c$ this case is most likely to have no observational consequence. Another possible case arises when $\sigma$ is still in slow-roll (or close to critically damped) in the entire observable Universe at the end of inflation. In this case, $\sigma$ obtains a statistically homogeneous spectrum of superhorizon perturbations. The cosmological consequences of such an extra light field have been extensively studied in the literature (see for example \cite{Lyth:2009zz}), and hence we do not consider it here.}. Thanks to the survival of this large value until the end of inflation, it becomes feasible to conjecture that $\sigma$ leaves some sort of observable imprint. In the following, we refer to those spatial regions where $\sigma\gtrsim\sigma_c$ at the end of inflation as \emph{out-of-equilibrium patches} \cite{Sanchez:2014cya}. Since we are interested in situations where out-of-equilibrium patches only occupy a small fraction of the observable Universe, the field configuration in those regions can be considered as an \textit{out-of-equilibrium remnant} from the primary epoch.

The feasibility of finding $\sigma$ in the interphase between the slow-roll and the oscillatory regime at the end of inflation is discussed in detail in Sec.~\ref{sec5}. For now, we implicitly assume the necessary parameter tuning so that this is indeed the case. Setting aside this question, the consistency of the above scenario already imposes the following important constraints. To secure that the energy density of $\sigma$ remains subdominant during inflation we must impose $\rho_\sigma\ll H_*^2m_P^2$ at the onset of slow-roll inflation, which is when $\sigma$ obtains its largest value $\sigma_{\rm sr}$. Using also Eq.~(\ref{eq72}), this condition translates into
\[\label{eq60}
\sigma_*<c_\sigma^{-1/2}\exp\left(c_\sigma N_{\rm sr}^p/3\right)m_P\,,
\]
where $m_P$ is the reduced Planck mass. Moreover, to have a chance of finding the field with a relatively large value $\sigma\sim\sigma_c$ in sparse regions of the Universe, we must enforce the condition $\sigma_c<\sigma_*$. Combining this with Eq.~(\ref{eq60}) and writing $\sigma_c\sim g^{-1}H_*$ we obtain
\[\label{eq17}
g\gg c_\sigma^{1/2}\exp\left(c_\sigma N_{\rm sr}^p/3\right)\frac{H_*}{m_P}\,.
\]
Imposing now that $g\leq{\cal O}(1)$, the existence of allowed values for $g$ demands that
\[\label{eq44}
N_{\rm sr}^p\ll\frac3{c_\sigma}\ln\frac{m_P}{c_\sigma^{1/2}H_*}\,.
\]
To estimate the upper bound we use $H_*/m_P<3.6\times10^{-5}$ \cite{Ade:2015lrj} and $c_\sigma={\cal O}(10^{-1})$, obtaining $N_{\rm sr}^p\ll 400$. This affords us to consider an epoch of primary inflation lasting for a few tens of $e$-foldings at most. This is an important point to emphasize, for it shows that \emph{primary inflation is not an unconstrained epoch} in our framework. To put it differently, the mechanisms considered in Sec.~\ref{sec14} can affect CMB temperature fluctuations on large scales only if primary slow-roll inflation is relatively short-lived. We stress that this conclusion lies along the same line of the findings in \cite{Wetterich:2015gya}, where the author shows that initial conditions at the beginning of inflation may affect the spectrum of cosmic fluctuations if the primary phase is not too large (see also \cite{Chen:2013eaa}).
 
To close this section, we remark that if $\sigma$ is in the equilibrium state in de Sitter space (as expected after a sufficiently prolonged phase of slow-roll inflation), typical expectation values are of order $\sigma\sim c_\sigma^{-1}H/2\pi$, at most \cite{Starobinsky:1994bd}. In that case, the condition $\sigma_c<\sigma_*$ translates into $g>2\pi c_\sigma$, which becomes incompatible with $g\leq{\cal O}(1)$ in our range of interest \mbox{$c_\sigma={\cal O}(10^{-1})$}. Therefore, the scenario considered here demands that existence of a sustained phase of fast-roll inflation to generate the necessary condition $\sigma_c<\sigma_*$. 

\section{Stochastic distribution of out-of-equilibrium remnants}\label{sec10}
Since we envisage the large-angle anomalies as the consequence of the out-of-equilibrium patches developed by isocurvature fields at the end of inflation, we need to describe the main properties of their stochastic distribution. Then, for a single isocurvature field $\sigma$ we must keep track of its associated probability density carrying the information on field correlations \emph{only} in the range of scales where CMB anomalies arise. To do so, we need to depart from the usual stochastic approach to inflation, for in that case the resulting probability density, while dictated by the Fokker-Planck equation in Eq.~(\ref{eq15}), contains the information on field correlations \textit{on all scales} that are superhorizon at the end of inflation.

\subsection{A modified Fokker-Planck equation}\label{sec9}
To trace field correlations on a given range of scales only we must follow the stochastic evolution of the classical field configuration built as superposition of the modes in the range of interest. As discussed in \cite{Sanchez:2014cya}, the simplest manner to carry out this filtering is by switching off the diffusion coefficient in Eq.~(\ref{eq15}) once the shortest scales of interest have exited the horizon. Therefore, we consider the scale-dependent diffusion coefficient
\[\label{eq58}
{\cal D}_k\equiv{\cal D}\,\theta(t_k-t)\,,
\]
where $\theta$ is the step function \cite{Abramowitz} and $t_k$ is the time of horizon exit for modes with comoving wavenumber $k$, i.e. $k=a(t_k)H$. This filtering of modes should result in a probability density not substantially different from the one obtained after smoothing the classical field at the end of inflation on the comoving scale $k^{-1}$. This expectation is based on the fact that both the scale-dependent filtering in Eq.~(\ref{eq58}) and the smoothing of the field remove structure on scales smaller than $k^{-1}$ while leaving unaffected the structure on larger scales. In our case, the advantage of using the scale-dependent filtering is that it provides us with a simple manner to keep track of the information of interest to us. The modified Fokker-Planck equation, obtained after the replacement ${\cal D}\to{\cal D}_k$ in Eq.~(\ref{eq15}), describes the evolution of the probability density $P_k(\sigma,t)$ associated to the classical field configuration with field correlations imprinted on all scales exiting the horizon before $t=t_k$.

As initial condition to solve for $P_k(\sigma,t)$, which we set when the largest cosmological scales exit the horizon $N_*$ $e$-foldings before the end of inflation, we impose
\[\label{eq43}
P_k(\sigma,t_*)=\delta(\sigma-\sigma_*)\,.
\]
We remark that this condition may be argued to be in conflict with Eq.~(\ref{eq38}), imposed at the onset of the slow-roll phase. The reason is that if the field peaks at $\sigma=\sigma_{\rm sr}$ at the onset of the slow-roll, inflationary fluctuations increase the field variance to $\Sigma^2\simeq(H^2/4\pi^2)N_{\rm sr}^p$ by the time of horizon crossing for the largest cosmological scales. Since we consider a non-negligible $N_{\rm sr}^p$, $\Sigma^2$ has a finite value at $t=t_*$. This is why the condition in Eq.~(\ref{eq43}) might be criticized as problematic, or even wrong, when confronted with the condition in Eq.~(\ref{eq38}). Nevertheless, one has the right to impose Eq.~(\ref{eq43}) in the understanding that, in that case, $P_k(\sigma,t)$ does not contain any information on field correlations on comoving scales beyond ${\cal H}_*^{-1}$. For our purposes this does not represent a problem, for we are only interested in the range of scales probed in the CMB. Therefore, we impose the initial condition Eq.~(\ref{eq43}) and the boundary condition Eq.~(\ref{eq24}) to solve for Eq.~(\ref{eq38}). It is then straightforward to find that the solution to the modified Fokker-Planck equation is the Gaussian 
\[\label{eq29}
P_k(\sigma,t)=\frac1{\sqrt{2\pi\Sigma^2_k(t)}}\,\exp\left[-\frac{(\sigma-\bar\sigma)^2}{2\Sigma^2_k(t)}\right]\,,
\]
where the mean field $\bar\sigma$ and the variance $\Sigma_k^2$ (not to be confused with $\Sigma^2$ in Eq.~(\ref{eq10})) are
\[\label{eq30}
\bar\sigma(t)=\sigma_*e^{-c_\sigma N/3}\quad,\quad\Sigma^2_k(t)=\frac{3H^2}{8\pi^2c_\sigma}
\left(1-e^{-\frac{2c_\sigma}3 H(t_k-t_*)}\right)e^{-\frac{2c_\sigma}3H(t-t_k)}\,,
\]
where now $N=H(t-t_*)$ counts the number of $e$-foldings elapsed after the largest cosmological scales exit the horizon.

For $t<t_k$, the evolution of the probability density is indistinguishable from the one obtained in the standard approach to stochastic inflation, in which $\Sigma^2_k\propto N$. For $t>t_k$, the behavior of the variance is very different. Owing to the scale-dependent filtering, the classical field configuration is no longer sourced by the continuous outflow of modes. As a result, $\Sigma_k$ decreases exponentially in the timescale $(3/c_\sigma)H^{-1}$ due to the curvature of the scalar potential $V(\sigma)$. This implies that $\Sigma_k$ can become significantly reduced at the end of inflation if $c_\sigma$ is not sufficiently small. In turn, as discussed in Sec.~\ref{sec5}, too small a value for $\Sigma_k$ at the end of inflation can increase significantly the parameter tuning necessary for the emergence of out-of-equilibrium patches. Writing $t_k-t_*=H_*^{-1}\log (k/{\cal H}_*)$, where ${\cal H}_*$ is the comoving scale crossing the horizon at $t=t_*$, we evaluate Eq.~(\ref{eq30}) at the end of inflation
\[\label{eq59}
\bar\sigma(t_{\rm end})=\sigma_*e^{-c_\sigma N_*/3}\quad,\quad\Sigma^2_k(t_{\rm end})=\frac{3H^2}{8\pi^2 c_\sigma}\,e^{-\frac{2c_\sigma}3N_*}
\left[\left(\frac{k}{{\cal H}_*}\right)^{\frac{2c_\sigma}3}-1\right]\,,
\]
where $\Sigma_k^2(t_{\rm end})$ inherits the scale-dependence of the diffusion coefficient ${\cal D}_k$. To recover the scale-independent result in Eq.~(\ref{eq36}) it suffices to consider the limit $t_k\to t_{\rm end}$, for which \mbox{$(k/{\cal H}_*)=e^{N_*}$}, thus allowing all superhorizon modes to contribute to the field variance.

We remark that to obtain the evolution of the probability density we have employed the boundary condition in Eq.~(\ref{eq24}), usually considered in the standard approach to stochastic inflation. This approach, however, must be modified due to the particle production mechanism operating for $\sigma\lesssim\sigma_c$. Below we impose absorbing barrier boundary conditions (see Appendix \ref{sec2}) to obtain an approximation to the stochastic field dynamics.

\subsection{Abundance of remnants}\label{sec5}
Despite its drawbacks, the usefulness of the boundary condition in Eq.~(\ref{eq26}) is that it provides us with a simple analytical estimate of the fraction of the probability density $P_k(\sigma,t)$ above the barrier at $\sigma_c$, where field interactions are still negligible. This fraction is obtained after integrating $P_k(\sigma,t_{\rm end})$ in the region $\sigma\geq\sigma_c$. Using Eq.~(\ref{eq59}) we find \cite{Sanchez:2014cya}
\[\label{eq3}
{\cal F}(k)=\int_{\sigma_c}^\infty \!\!P_k(\sigma,t_{\rm end})\,d\sigma=\frac12+\frac12\,\textrm{Erf}\left[\xi(k,t_{\rm end})\right]\,,
\]
where $\xi(k,t)\equiv\frac{\bar\sigma(t)-\sigma_c}{\sqrt2\,\Sigma_k(t)}$. The above represents the expected fraction of inflated volume where field correlations can be found on comoving scales ranging from $k^{-1}$ to ${\cal H}_*^{-1}$. We recall that the upper bound ${\cal H}_*^{-1}$ is set by the initial condition Eq.~(\ref{eq43}).

\begin{figure}[htbp]
\centering\epsfig{file=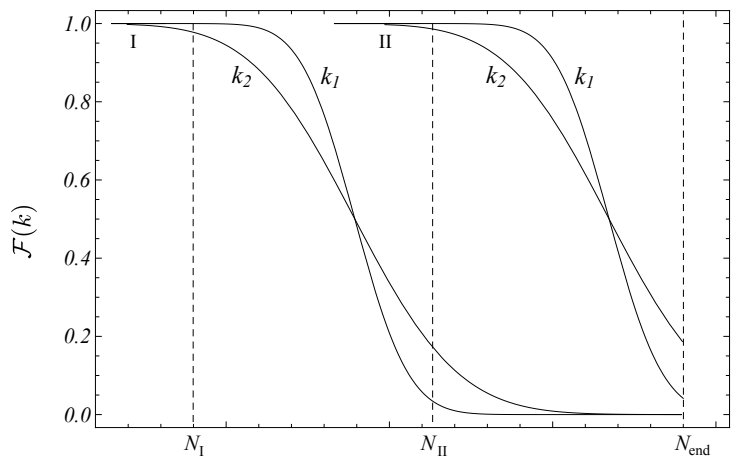,width=9.5cm}
\caption{Plot of the expected fraction ${\cal F}(k)$ for different scales, with $k_1<k_2$, and in two different instances, labeled I and II, attending to the field value $\sigma_{\rm sr}$ at the onset of slow-roll inflation.}\label{fig8}
\end{figure}

In Fig.~\ref{fig8} we schematically illustrate a key aspect in our framework. Namely, that tuning appropriately the model parameters it is possible to arrange the transition of $\sigma$ to the oscillatory regime at any time during inflation. To show this, we depict the expected fraction ${\cal F}(k)$ for different comoving scales $k_1<k_2$ as a function of the number of $e$-foldings, $N$, in two different situations, labeled I and II. In case I, the value $\sigma_{\rm sr}$ is chosen small so that the transition to the oscillatory regime takes place around $N_{\rm I}$, well before the end of inflation at $N_{\rm end}$. As a result, at the end of inflation ${\cal F}\simeq0$ in all scales of interest, thus implying the absence of out-of-equilibrium patches in the observable Universe. In case II, $\sigma_{\rm sr}$ is chosen so that the transition to the oscillatory regime happens at a later time $N_{\rm II}>N_{\rm I}$. In this case, out-of-equilibrium patches are expected to emerge at the end of inflation with abundances determined by the fractions ${\cal F}(k_2)>{\cal F}(k_1)$. Since ${\cal F}<1$ in the case shown (in particular ${\cal F}(k_1)\ll1$), case II describes the emergence of out-of-equilibrium patches in sparse regions of the observable Universe. Larger values of $\sigma_{\rm sr}$ displace the plotted curves to the right, which then results in an increase of ${\cal F}(k)$ on all scales. Consequently, we obtain a distribution of larger out-of-equilibrium patches covering a larger fraction of the observable Universe. Yet another possible case, not shown in the plot, is when $\sigma_{\rm sr}$ is sufficiently large so that ${\cal F}\simeq1$ on all scales of interest at the end of inflation. In this case, the entire observable Universe can be considered as an out-of-equilibrium patch where $\sigma$ remains oblivious of its interaction with $\chi$ during inflation.

An important issue regarding out-of-equilibrium patches is that of their shape. Recalling their definition, out-of-equilibrium patches correspond to the regions where a Gaussian random field, $\sigma$ in our case, is above the threshold $\sigma=\sigma_c$. Although these regions have a complicated structure \cite{adler}, it can be shown that the triaxial ellipsoid approximation is a valid description in the immediate neighborhood of the peak, and that high peaks tend to be more spherically symmetric than lower ones. In turn, nearly spherical shapes only emerge when very large thresholds (i.e. rarely occurring peaks) are considered \cite{Bardeen:1985tr} (see also \cite{Marcos-Caballero:2015lxp} for an application to the study of CMB peaks).

Another important issue is related to the likelyhood of the scenario considered here. To assess whether the emergence of patches is a probable outcome we need to compute the fraction of the field distribution that results in the emergence of patches at the end of inflation. Similarly to the fraction in Eq.~(\ref{eq3}), this is given by the integral
\[\label{eq74}
I\equiv\int_{\Delta\sigma}G\,d\sigma\,,
\]
where $\Delta\sigma$ represents the range of $\sigma$ (taken at the end of the fast-roll) leading to the formation of out-of-equilibrium patches and $G$ is a Gaussian distribution with zero mean and variance $\Sigma^2$ given by Eq.~(\ref{eq10}). Using the absorbing barrier approximation and assuming the appropriate conditions for the emergence of patches we obtain (see Appendix \ref{sec12})
\[\label{eq33}
I={\rm Erf}\left(\frac{\sigma_{\rm sr}(\xi_{\rm max})}{\sqrt2\,\Sigma}\right)-{\rm Erf}\left(\frac1{\sqrt2}\right)\,,
\]
where 
\[\label{eq46}
\frac{\sigma_{\rm sr}(\xi_{\rm max})}{\Sigma}=\frac{1+\sqrt2\left(\Sigma_k(t_{\rm end})/\sigma_c\right)\,\xi_{\rm max}}{1+\sqrt2\left(\Sigma_k(t_{\rm end})/\sigma_c\right)\,\xi_{\rm min}}\,.
\]

\begin{figure}[htbp]
\epsfig{file=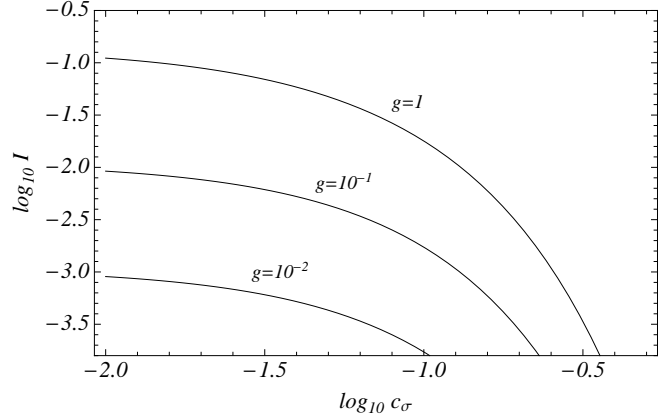,width=8.5cm}
\centering\caption{Plot of the integral $I\equiv\int_{\Delta\sigma}G\,d\sigma$ as a function of $c_\sigma$ and for different values of $g$.}\label{fig13}
\end{figure}

Our results are plotted in Fig.~\ref{fig13}, where we evaluate Eq.~(\ref{eq33}) for different values of the coupling $g$, as indicated. To build the plot we demand that out-of-equilibrium patches with typical sizes corresponding to all CMB scales emerge in 1-20\% of the observable Universe, namely $10^{-2}\leq {\cal F}(k)\leq0.2$ with $k=e^{N_{\rm CMB}}{\cal H}_*$. As $g$ decreases, the emergence of patches becomes increasingly unlikely. At $g$ fixed, since the variance $\Sigma_k^2(t_{\rm end})$ becomes weakly dependent on $c_\sigma$ in the range $c_\sigma\leq{\cal O}(10^{-1})$, a change in $c_\sigma$ entails only a moderate variation in $I$. For $c_\sigma>{\cal O}(10^{-1})$, the variance $\Sigma_k^2(t_{\rm end})$ decreases rapidly with $c_\sigma$, and so does $I$ as a result. Our plot clearly reveals a small value of $I$ for natural values of $g$ and with $c_\sigma={\cal O}(10^{-1})$, and hence a preference for case III in Fig.~\ref{fig17} (see Appendix \ref{sec12}). Therefore, 
the initial value $\sigma_{\rm sr}$ must be significantly tuned so that out-of-equilibrium patches can emerge at the end of inflation. Nevertheless, in Sec.~\ref{sec7} we go beyond the absorbing barrier approximation and show that such a tuning can be much alleviated. 

\subsection{Scale-dependent distribution}\label{sec15}
Here we follow \cite{Sanchez:2014cya} to review the scale-dependent behavior of the distribution of patches. From the interpretation of the fraction in Eq.~(\ref{eq3}), it follows that the differential ${\cal F}'(k)\,dk$ gives the fraction of the inflated volume with field correlations on scales in the interval $[k+dk,k]$. The volume of the observable Universe that corresponds to this fraction is \mbox{$dV_k={\cal H}_*^{-3}\times{\cal F}'(k)\,dk$}. To obtain a simplified description of the distribution of out-of-equilibrium patches, we will focus on the spatial regions where $\sigma\geq\sigma_c$ is correlated on the comoving scale $k^{-1}$. For the sake of brevity, we refer to these regions as \emph{$k$-patches}. Since out-of-equilibrium patches correspond to high peaks of Gaussian random fields, then their average shape tends to be spherically symmetric \cite{Bardeen:1985tr}. In that case, we can assume that the typical comoving volume occupied by a $k$-patch is of order $k^{-3}$. This estimate allows us to compute the typical number of $k$-patches witin the observable Universe, given by $dN_k=k^3dV_k$, along with its number density $n(k)$ per unit interval $dk$
\[\label{eq4}
n(k)\equiv \frac{dN_k}{dk}={\cal F}'(k)\left(\frac{k}{{\cal H}_*}\right)^3\,.
\]
In general, the average shape of out-of-equilibrium patches can deviate from a sphere, and hence the actual magnitude of $n(k)$ and its scale-dependence will also deviate from those obtained in Eq.~(\ref{eq4}). Nevertheless, the latter can be expected to become a reasonable approximation when out-of-equilibrium patches correspond to high peaks of Gaussian fields. Having this caveat in mind, in the following we use Eq.~(\ref{eq4}) to obtain qualitative features of the distribution of patches.

Now, to estimate the number density of $k$-patches in the last scattering surface, we need to compute the probability that a $k$-patch intersects it, which we denote by $P_{\rm lss}(k)$. To do so, we take the observable Universe to be a box of comoving size $2{\cal H}_*^{-1}$ and $k$-patches to be spheres of comoving radius $k^{-1}/2$ with the center randomly located. In that case, $P_{\rm lss}(k)$ can be approximated by (see Appendix \ref{app2})
\[\label{eq83}
P_{\rm lss}(k)=\frac{\pi}{2}\frac{{\cal H}_*}{k}\left[1+\frac1{12}\left(\frac{{\cal H}_*}{k}\right)^2\right]\,.
\]
If we further assume that the typical scale of the resulting intersection is of order $k^{-1}$, the number density of $k$-patches (per unit interval $dk$) in the last scattering surface is simply
\[\label{eq6}
{\cal N}(k)\equiv n(k)\,P_{{\rm lss}}(k)\simeq\frac{\pi}2\,{\cal F}'(k)\left(\frac{k}{{\cal H}_*}\right)^2\,.
\]

\begin{figure}[htbp]
\centering\epsfig{file=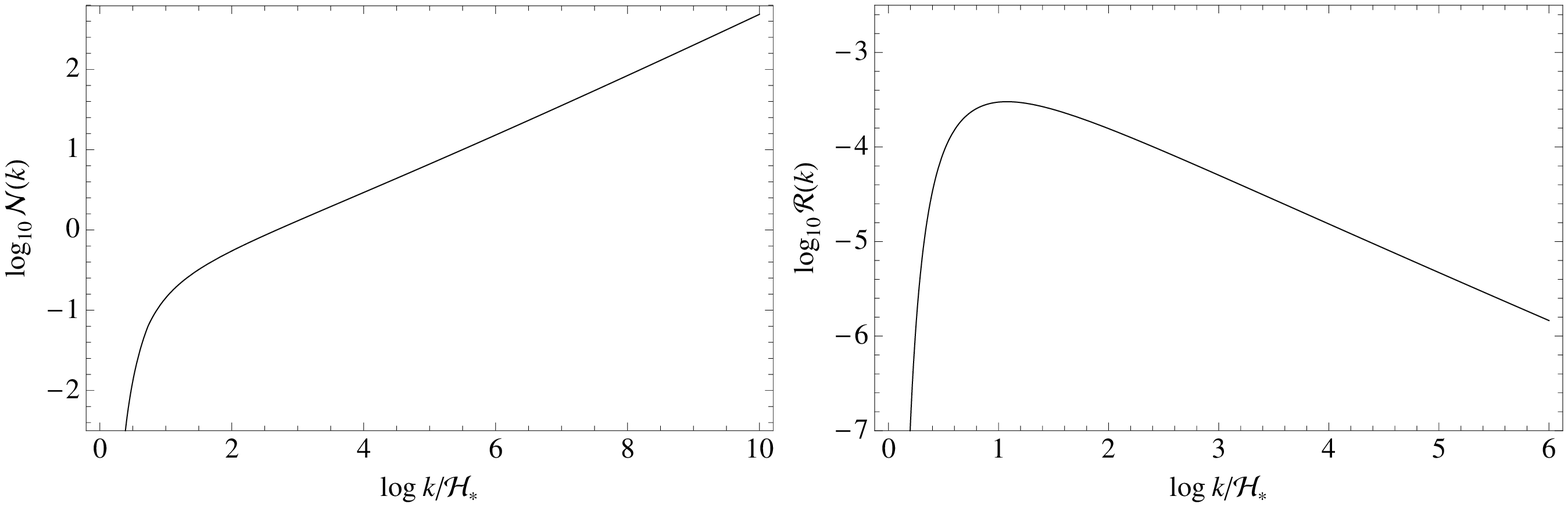,width=15cm}\caption{Number density ${\cal N}(k)$ (left-hand panel) and relative number density ${\cal R}(k)$ (righthand panel) of $k$-patches in last scattering surface. In the case shown \mbox{$c_\sigma=0.10$}, $g=0.5$ and $g\sigma_*\simeq23H_*$.}\label{fig2}
\end{figure}

In Fig.~\ref{fig2} (left-hand panel) we plot the predicted ${\cal N}(k)$ taking \mbox{$c_\sigma=0.10$}, $g=0.5$ and $g\sigma_*\simeq23H_*$. For comoving wavenumbers approaching ${\cal H}_*$, the number density ${\cal N}$ goes to zero. This behavior is the result of imposing the initial condition in Eq.~(\ref{eq43}) and the boundary condition in Eq.~(\ref{eq26}). On the one hand, the initial condition in Eq.~(\ref{eq43}) and the scale-dependent diffusion coefficient ${\cal D}_k$ result in a $\delta$-like distribution when dealing with correlations on the largest scales, i.e. $P_{k={\cal H}_*}(\sigma,t)=\delta(\sigma-\bar\sigma,t)$. On the other hand, the use of an absorbing barrier implies that a $\delta$-like distribution can only be either above or below $\sigma_c$. Therefore, when $\bar\sigma=\sigma_c$, the fraction ${\cal F}(k={\cal H}_*)$ passes from 1 to zero discontinuously, thus explaining the abrupt fall to zero in Fig.~\ref{fig2} as $k$ approaches ${\cal H}_*$.

Our plot also shows a growing number density of patches on smaller scales. This result is expected, for the continuous imprint of structure in the classical field amounts to the growth of the field variance. In turn, this gives the field a greater chance to be above $\sigma_c$ in patches of smaller size. As a result, out-of-equilibrium patches become more abundant on smaller scales than on larger ones. This growing number entails potentially harmful consequences: if some mechanism is provided whereby out-of-equilibrium patches come to affect CMB temperature anisotropies on large-angular scales, then this effect should be more noticeable on small scales, where out-of-equilibrium patches are more abundant. However, there seems to be no indication of an anomalous or non-Gaussian spectrum on such scales, apart from the persistence of a power asymmetry extending to $\ell=600$ \cite{Ade:2013ydc}. Consequently, our framework must provide an explanation for the non-detection of irregularities on smaller scales.

To assess the implications of the out-of-equilibrium patches for the CMB temperature anisotropies we find the abundance of $k$-patches relative to the total number density of patches of size $k^{-1}$ contained in the last scattering surface, which is given by $n_{\rm lss}(k)\propto k^2$. The relative number density of $k$-patches per unit interval $dk$ in the last scattering surface is
\[\label{eq25b}
{\cal R}(k)\equiv \frac{{\cal N}(k)}{n_{\rm lss}(k)}\simeq\frac{\pi}{2^7}\,{\cal F}'(k)\,.
\]
This ratio is depicted in the righthand panel of Fig.~\ref{fig2}, where the curve shown corresponds to the same parameters as in the left-hand panel. Since ${\cal R}(k)\propto {\cal F}'(k)$, the ratio vanishes in the limit $k\to{\cal H}_*$. As previously explained, this behavior is the result of boundary and initial conditions. Using Eqs.~(\ref{eq59}) and (\ref{eq3}), we can compute the behavior of ${\cal R}(k)$ for larger $k$, obtaining
\[\label{eq61}
{\cal R}(k)\propto \left(\frac{k}{{\cal H}_*}\right)^{-1+2c_\sigma/3}\ln^{-3/2}\left(\frac{k}{{\cal H}_*}\right)\,,
\]
which gives a decreasing function of $k$. Therefore, despite the increasing number of out-of-equilibrium patches on smaller scales, their relative number quickly decreases. As a result, out-of-equilibrium patches are outnumbered by adiabatic ones, where the inflaton imposes its nearly scale-invariant, Gaussian perturbation. Consequently, one can expect that the perturbation spectrum on smaller scales is dominated by the inflaton field. Owing to this, our scenario enjoys the appropriate qualitative behavior to make it compatible with both the generation of sizable effects on the CMB on large-angular scales, provided the appropriate mechanism is considered, and the absence of an observable effect on smaller scales. 

Fig.~\ref{fig2} also shows that ${\cal R}(k)$ peaks at a given scale. This is where the significance of any anomalous signature imprinted in the CMB (seeded by out-of-equilibrium patches) is at its highest, and hence it constitutes a sort of preferred scale. As it stands, the existence of such a scale is a consequence of initial and boundary conditions imposed on the field distribution. Consequently, one can expect that the scale-dependence computed in Eq.~(\ref{eq61}) will also apply to scales $k\ll{\cal H}_*$. In this regard, in the next section we see how the production of superhorizon fluctuations of $\chi$ can prevent the appearance of a scale maximizing ${\cal R}(k)$. Nevertheless, we can anticipate that in the limit of rapid growth of $m_\sigma$, due for example to either a large coupling $g$ or a large multiplicity for the $\chi$ field, the transition to the oscillatory phase can take place in much less than a Hubble time, and hence we should recover the results from the absorbing barrier. Therefore, we emphasize that, even after dispensing with the boundary conditions in Eq.~(\ref{eq26}), a preferred scale might still arise.

\subsection{Beyond the absorbing barrier}\label{sec7}
As discussed in Appendix \ref{sec2}, although the absorbing barrier approximation provides us with a simple estimate of the volume of observable Universe covered by out-of-equilibrium patches at the end of inflation, it fails to reproduce important features of the physical system. Most importantly, this approximation may be inappropriate, for it assigns a vanishing expectation value to the field as soon as this reaches the absorbing barrier, thus entailing its instantaneous disappearance. In a more realistic situation, the transition to the oscillatory stage is expected to occur in the Hubble timescale, for the latter is the relevant one for the production of inflationary fluctuations. Moreover, the absorbing barrier approximation falls short too when it comes to estimate parameter tuning.  As shown below, the estimate in Eq.~(\ref{eq33}) turns out to be a too pessimistic one when we dispense with the condition in Eq.~(\ref{eq26}). To ease these difficulties, we consider a phenomenological model that takes into account the finite time required for the field to enter its oscillatory stage. In this new approximation, we write the probability density as
\[
P_k^{\rm (ext)}(\sigma,t)=\theta(\sigma-\sigma_c)P_k(\sigma,t)+P_k^{\rm (ph)}(\sigma,t)\,,
\]
where $P_k(\sigma,t)$ is given by Eq.~(\ref{eq29}) and the phenomenological part $P_k^{\rm (ph)}(\sigma,t)$ (derived in Appendix \ref{sec16}) accounts for the gradual depopulation of the slow-roll phase in the timescale $\tau_t$, which we take to be of order $H^{-1}$.

\begin{figure}[htbp]
\centering\epsfig{file=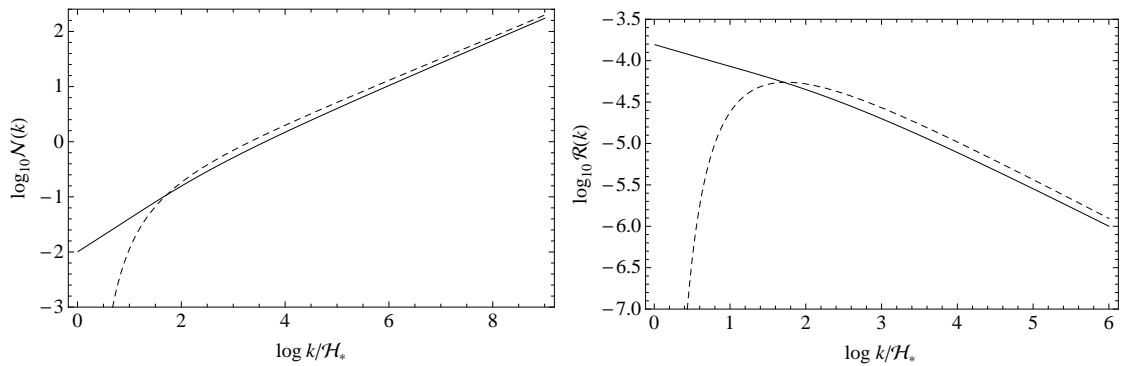,width=15cm}\caption{Number density ${\cal N}(k)$ (left-hand panel) and relative number density ${\cal R}(k)$ (righthand panel) of $k$-patches as obtained following the phenomenological approach (solid line). For comparison we include the prediction in the case of an absorbing barrier (dashed line). The case shown corresponds to \mbox{$c_\sigma=0.05$}, $g=0.7$, $\tau_t=H_*^{-1}$ and $N_*=60$.}\label{fig9abc}
\end{figure}

Similarly to the absorbing barrier case, we find the fraction of the inflated volume in out-of-equilibrium patches by integrating the above probability\footnote{In contrast to the case of an absorbing barrier, this fraction does not admit an analytical expression and numerical integration becomes necessary to evaluate it.}
\[\label{eq64}
{\cal F}_{\rm ext}(k)=\int_0^\infty P_k^{\rm (ext)}(\sigma,t)\,d\sigma=\int_0^{\sigma_c}P_k^{\rm (ph)}(\sigma,t)\,d\sigma+{\cal F}(k)\,.
\]
Using this, we compare now the scale-dependent distribution of patches obtained following the two approaches. To do so, we compute ${\cal F}'_{\rm ext}(k)$ numerically to obtain the ``extended'' version of ${\cal N}(k)$ and ${\cal R}(k)$, following Eqs.~(\ref{eq6}) and (\ref{eq25b}). To perform a meaningful comparison between the two approaches, we tune the model parameters to obtain an equal abundance of patches at the end of inflation in both cases, i.e. ${\cal F}(k)={\cal F}_{\rm ext}(k)$. This can be achieved, for example, by choosing the appropriate $\sigma_{\rm sr}$ in each case. In Fig.~\ref{fig9abc} we plot the prediction for ${\cal N}(k)$ and ${\cal R}(k)$, as obtained using the absorbing barrier approximation (dashed line) and the phenomenological approach (solid line). The case shown corresponds $g=0.7$, $c_\sigma=0.05$, $\tau_t=H_*^{-1}$ and $N_*=60$. In both cases, out-of-equilibrium patches arise with an abundance ${\cal F}=0.15$. The results for both approaches differ in a number of aspects. Firstly, in the phenomenological approach ${\cal N}(k)$ remains finite in the limit $k\to{\cal H}_*$. This happens because the amplitude of the $\delta$-like distribution $P_{k={\cal H}_*}(\sigma,t)$ decreases exponentially in the timescale $\tau_t$ after crossing the barrier, instead of vanishing as in the absorbing barrier approximation. As a result, integrating $P_k^{\rm (ph)}$ returns a finite result. Furthermore, owing to the persistence of out-of-equilibrium patches, in particular those on the largest scales, there is no wavenumber \mbox{$k<{\cal H}_*$} maximizing ${\cal R}(k)$. Nevertheless, this sort of preferred scale can still appear if \mbox{$H_*\tau_t<1$}, since the phenomenological model resembles an absorbing in the limit\footnote{This might be the case when the $\chi$ field has a large multiplicity or for relatively large couplings $g$.} $\tau_t\to0$.

We wish to remind the reader that the approximations here considered are not meant to provide an accurate computation of the distribution of out-of-equilibrium patches, but an educated estimate of their abundance and expected scale-dependent behavior. An accurate determination of the distribution of out-of-equilibrium patches, on the other hand, requires numerical simulations, which is beyond the scope of this paper. Bearing this caveat in mind, we manage to show that it becomes relatively easy to find model parameters, both for an absorbing barrier and for our phenomenological model, so that out-of-equilibrium patches emerge with an abundance in the right ballpark and a scale-dependent behavior well suited, at least in principle, to become the seeds for the large-angle anomalies observed in the CMB.

\begin{figure}[htbp]
\centering\epsfig{file=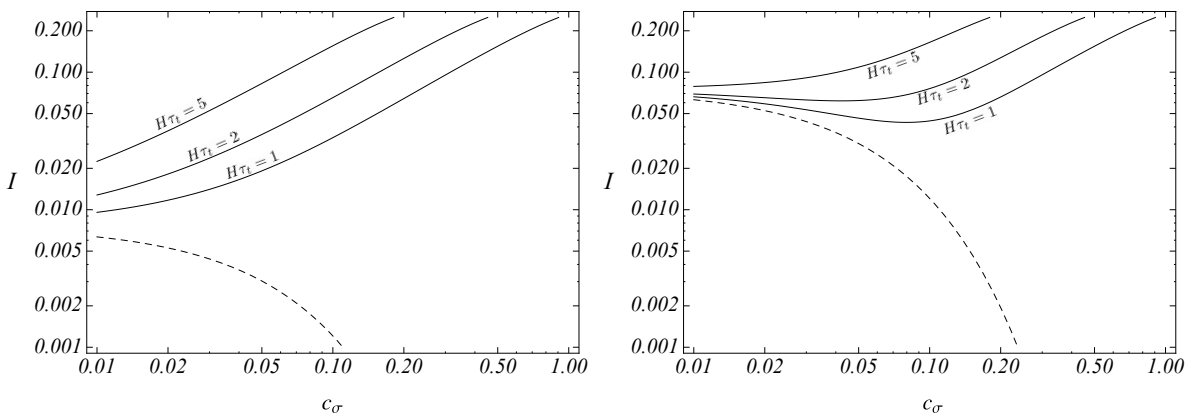,width=15.5cm}\caption{Predicted curves for the integral $I$ in terms of $c_\sigma$ and for several values of $\tau_t$, as indicated. We use $g=0.1$ (left-hand panel) and $g=1$ (righthand panel). The dashed line corresponds to the case of an absorbing barrier.}\label{fig15}
\end{figure}

Next, we summarize the results concerning the level of parameter tuning. Similarly to the absorbing barrier case, the integral $I$ is given by Eq.~(\ref{eq33}), but due to the finite transtion time $\tau_t$ we obtain now (see Appendix \ref{sec11})
\[\label{eq41}
\frac{\sigma_{\rm sr}(\xi_{\rm max})}{\Sigma}=\exp\left(\frac{\sqrt2\,\Sigma_k(t_{\rm end})\Delta\xi}{\sigma_c}+\frac23\,c_\sigma H_*\tau_t\right)\,.
\]
In Fig.~\ref{fig15} we show our results using $H_*\tau_t=1,2,5$. We take $g=0.1$ in the left-hand panel and $g=1$ in the righthand one. Similarly to the case shown in Fig.~\ref{fig13}, in both cases we demand the emergence of patches on all CMB scales with an expected abundance \mbox{$10^{-2}\leq{\cal F}_{\rm ext}(k_{\rm CMB})\leq0.20$}, which gives $\Delta\xi=\xi_{\rm max}-\xi_{\rm min}\simeq1$. For comparison, we include the prediction for an absorbing barrier (dashed line). For sufficiently small $c_\sigma$ (depending on $g$) and $H_*\tau_t={\cal O}(1)$, no significant differences arise between the two approaches. This is because for small $c_\sigma$ the field distribution performs a slow-roll motion. Since the introduction of $\tau_t$ has a subdominant effect in this case ($\widetilde{\Delta t}\simeq \Delta t$ in Eq.~(\ref{eq85})), the predicted $I$ approaches the result obtained for an absorbing barrier. For $c_\sigma$ large enough, the introduction of $\tau_t$ becomes the dominant effect ($\widetilde{\Delta t}\simeq2\tau_t$ in Eq.~(\ref{eq85})) and the behavior of $I$ changes, starting to grow with increasing $c_\sigma$ in contrast to the absorbing barrier case. We find that even a conservative departure from the absorbing barrier case, like $H_*\tau_t=1$, can increase $I$ substantially, thus alleviating the fine-tuning problem revealed in Fig.~\ref{fig13}. 

A remarkable feature of our results is that $I$ becomes independent of $g$ above certain threshold for $c_\sigma$. This happens because the exponential in Eq.~(\ref{eq41}) becomes independent of $g$ when the term in $\tau_t$ becomes the dominant one. Note that the first term in the exponential grows with $g$ due to its inverse dependence with $\sigma_c\propto g^{-1}$. Since $g\leq1$, the left-hand side in Eq.~(\ref{eq41}) becomes independent of $g$ for $c_\sigma>\Sigma_k(t_{\rm end})/H_*$. Using Eq.~(\ref{eq59}) and expanding $\Sigma_k(t_{\rm end})$ to first order in $c_\sigma$, this conditions translates into
\[\label{eq79}
c_\sigma>\left(\frac{N_{\rm CMB}}{4\pi^2}\right)^{1/2}\exp(-c_\sigma N_*/3)\,,
\]
which is satisfied for $c_\sigma\geq{\cal O}(10^{-1})$. In that case, using $H_*\tau_t\geq{\cal O}(1)$ we obtain $I\geq{\cal O}(10^{-2})$, implying that the emergence of patches at the end of inflation is a relatively likely outcome, with a probability of the order of a few per cent. This is a very encouraging result, for it suggests that out-of-equilibrium patches may be feasible candidates to become the seeds of large-angle CMB anomalies, thus offering an avenue to account for the latter without having to invoke an alternative scenario more unlikely that the very existence of anomalies. Yet another feature worth stressing is that one can expect to obtain $H_*\tau_t\geq{\cal O}(1)$ without having to impose that the $\chi$ field belongs to large GUT groups \cite{Enomoto:2013mla}. On the contrary, the expectation when $\chi$ belongs to large GUT groups is that $H_*\tau_t\ll{\cal O}(1)$, in which case the emergence of patches becomes a very unlikely event, as Fig.~\ref{fig15} demonstrates.

\section{Implications for the Cosmic Microwave Background}\label{sec14}
Until now, we have investigated the generation of out-of-equilibrium patches in the observable Universe and have identified the necessary conditions so that their emergence becomes a likely event. Our goal in this section is to explore a number of mechanisms to determine if the emergence of out-of-equilibrium patches can affect temperature fluctuations in the CMB sufficiently to conjecture that their existence can be related to the large-angle CMB anomalies. In this sense, we emphasize that the overall purpose of this section is to assess the potential of the framework developed in previous sections to account for CMB anomalies. Consequently, the mechanisms examined below must be considered as a test of feasibility. To fully determine if any of the studied mechanisms provides an explanation preferred by data a dedicated analysis is necessary.

\subsection{The case of the Cold Spot}
The Cold Spot anomaly refers to a large, nearly circular region of the CMB sky, around \mbox{$\vartheta_{\rm cs}\simeq10^\circ$} in angular size in the southern hemisphere and with a significant temperature decrement. Since its first detection in 2004 \cite{Vielva:2003et}, the Cold Spot has been the subject of numerous statistical analysis (see \cite{Vielva:2010ng} for an extensive review). In order to explain this observation, a number of explanations have been considered in the literature: a local void \cite{Inoue:2006rd,Inoue:2006fn,Finelli:2014yha,Szapudi:2014eza,Szapudi:2014zha,Kovacs:2015hew} (see also \cite{Zibin:2014vaa,Nadathur:2014tfa,Marcos-Caballero:2015phy}), the Sunyaev-Zeldovich effect \cite{Cruz:2008sb}, the formation of a cosmic texture \cite{Cruz:2007pe}, multifield inflation \cite{Afshordi:2010wn}, or chaotic preheating \cite{Bond:2009xx} among others. Here we review the model proposed in \cite{Sanchez:2014cya}, in which the Cold Spot originates in the last scattering surface as the result of a local inhomogeneous reheating mechanism.

The idea underlying the scenario examined in \cite{Sanchez:2014cya} is that the inflaton's decay rate varies sufficiently in out-of-equilibrium patches to modify the curvature perturbation imprinted by the inflaton. On the contrary, in patches where $\sigma$ oscillates before the end of inflation, the inflaton decay rate remains unaffected since $\sigma\ll\sigma_c$. Consequently, the curvature perturbation does not become modified.

The first condition to impose is that out-of-equilibrium patches must have an expected number density ${\cal N}(k)={\cal O}(1)$ in the appropriate range of scales. To obtain a quick estimate of the comoving wavenumber $k_{\rm cs}$ corresponding to the Cold Spot, it suffices to assume a matter dominated Universe at present, obtaining \mbox{$\frac{k_{\rm cs}}{{\cal H}_*}\simeq\frac{\vartheta_{\rm dec}}{\vartheta_{\rm cs}}\,\Omega_m^{1/2}/(1+z_{\rm dec})^{1/2}\simeq3$}, where $\vartheta_{\rm dec}$ is the angle subtended by the horizon at the time of decoupling.  Figs.~\ref{fig2} and \ref{fig9abc} demonstrate the existence of model parameters to generate a sufficiently large abundance of patches in the appropriate scales. Moreover, as discussed in Sec.~\ref{sec15}, out-of-equilibrium patches become more abundant on smaller scales, which then should have observational implications. In turn, this feature might explain the presence of other anomalous spots, smaller than the Cold Spot, discovered in the CMB \cite{Vielva:2007kt,Ade:2013nlj}, although these are detected to a smaller significance.

\subsubsection{Local inhomogeneous reheating}
Assuming that our spectator field $\sigma$ modulates the decay rate of the inflaton, the contribution to the curvature perturbation due to inhomogeneous reheating is \cite{Dvali:2003em,Kofman:2003nx,Dvali:2003ar,Mazumdar:2003iy,Matarrese:2003tk,Tsujikawa:2003bn}
\[\label{eq28}
\zeta_\sigma=\alpha\left(\frac{\delta \Gamma}{\Gamma}\right)_{\rm dec}\,,
\]
where ``dec'' labels the time of inflaton decay and $\alpha$ is the efficiency parameter and, for simplicity, we assume that the isocurvature perturbation in $\sigma$ is completely converted into a curvature perturbation. Following \cite{Sanchez:2014cya}, we consider the case when the Universe becomes matter dominated after inflation and take the efficiency parameter $\alpha\simeq1/6$, thus assuming a late decay of the inflaton \cite{Dvali:2003em,Kofman:2003nx,Dvali:2003ar,Mazumdar:2003iy,Matarrese:2003tk,Tsujikawa:2003bn}.

Here we focus on two decay rates with different implications for the temperature fluctuations in the CMB. When $\Gamma(\sigma)$ is a growing function of $\sigma$, out-of-equilibrium patches result in an anticipated decay of the inflaton. As a result, the energy density undergoes an enhanced redshift in these patches, thus giving rise to enhanced underdensities at the time of decoupling. Since this results in a local enhancement of the temperature, a growing $\Gamma(\sigma)$ may offer an explanation for the anomalous hot spots detected in the CMB \cite{Vielva:2007kt,Ade:2013nlj}. This is the case, for example, of the decay rate \cite{Postma:2003jd}
\[\label{eq2b}
\Gamma(\sigma)=\Gamma_0\left[1+\left(\frac{\sigma}{M}\right)^q\right]^r\,,
\]
where $\Gamma_0$ is the unperturbed inflaton's decay rate, $q\geq1$, $M$ is a mass scale and $\sigma< M$ at the time of decay. In the opposite case, when $\Gamma(\sigma)$ is a decreasing function of $\sigma$, out-of-equilibrium regions result in a delayed decay of the inflaton, producing enhanced matter overdensities at the time of decoupling. Since an overdensity in the last scattering surface lowers the temperature, such a decay may be invoked to explain the Cold Spot. An example of this is the decay width resulting from the 2-body decay of the inflaton into $\psi$ particles \cite{Postma:2003jd}
\[\label{eq14}
\Gamma=\Gamma_0\left[1-\left(\frac{2m_\psi(\sigma)}{m_\phi}\right)^2\right]^{1/2}\,,
\]
where $m_\psi(\sigma)=\lambda\sigma$ and $\lambda$ is a dimensionless coupling. Since we take the Cold Spot as due to an enhanced overdensity in the last scattering surface, the decay in Eq.~(\ref{eq14}) will be the one of interest to us. Nevertheless, if we restrict ourselves to $q=2$ and $r=1$ and make the replacement $M\to\sqrt{qr}\lambda^{-1}m_\phi$ in Eq.~(\ref{eq2b}), we obtain from the latter the same contribution to the curvature perturbation as from Eq.~(\ref{eq14}). Therefore, given this identification, below we restrict ourselves to the decay rate in Eq.~(\ref{eq2b}).

The contribution to the curvature perturbation can then be written as \cite{Sanchez:2014cya}
\[\label{eq12}
\zeta_\sigma\simeq\alpha q r\left(\frac{\sigma_{\rm dec}}{M}\right)^q\left(\frac{\delta\sigma}{\sigma}\,\right)_{\rm end}\,.
\]
To compute the above, we use that \mbox{$\sigma_{\rm end}\simeq\sigma_c$} in out-of-equilibrium patches, but for $\sigma_{\rm dec}$ we need to specify the field evolution until the time of reheating. But before we do that, we note that in patches where $\sigma$ oscillates before the end of inflation, the typical value of $\sigma$ is determined by the amplitude of the oscillations (see Eq.~(\ref{eq80})). Therefore, in these patches $\sigma_{\rm dec}$ becomes suppressed in comparison to $\sigma_c$, which is the approximate field value in out-of-equilibrium patches. Since $\zeta_\sigma\propto \sigma_{\rm dec}^2$, the contribution to the curvature perturbation becomes suppressed by a factor $\sim\exp[-3N_{\rm osc}(\mbox{\boldmath$x$})]$, but depending on the evolution of $\sigma$ until reheating this suppression can become even more significant. In any case, since $\zeta_\sigma$ becomes unobservable in these patches, it does not become necessary to compute such a correction. To compute the fractional perturbation in Eq.~(\ref{eq12}) we proceed as follows. Using the spectrum in Eq.~(\ref{eq57}), the amplitude of a field perturbation at horizon crossing is $(\delta\sigma)_k\sim (H_*/2\pi)$. When $c_\sigma$ is not too small, this fluctuation can undergo a sizable evolution until the end of inflation. Since the evolution of $\sigma$ and its perturbation $\delta\sigma$ is determined by the same equation, the fractional perturbation $(\delta\sigma/\sigma)$ remains constant. Therefore, using Eq.~(\ref{eq72}) and $\sigma_{\rm end}\simeq\sigma_c$ we obtain
\[\label{eq11}
\left(\frac{\delta\sigma}{\sigma}\right)_{k,{\rm end}}\simeq \frac{H_*}{2\pi\sigma_c}\,\exp[-N(k)\,c_\sigma/3]\,,
\]
where $N(k)$ denotes the remaining number of $e$-foldings when the comoving wavenumber $k$ crosses outside the horizon. 

\subsubsection{Post-inflationary evolution}\label{sec4}
To study the superhorizon evolution of $\sigma$ after inflation we include a bare mass $m_0$ in the effective mass of $\sigma$, 
\[
m_\sigma^2=m_0^2+c_\sigma H^2+g^2\langle\chi^2\rangle\,.
\]
Although we implicitly assumed that $m_0$ is subdominant during inflation, now we allow for the possibility that $m_\sigma^2\simeq m_0^2$, for both $H^2$ and $\langle\chi^2\rangle$ decrease rapidly during the matter dominated epoch. Also, since $\sigma$ becomes negligible in those patches where it starts oscillating before the end of inflation, we are concerned only with the evolution of the classical field in out-of-equilibrium patches. In these, the interaction mass is subdominant, and hence the field equation can be approximated by
\[
\ddot\sigma+3H\dot\sigma+\left(m_0^2+c_\sigma H^2\right)\sigma\simeq0\,.
\]
Right after inflation, when $m_0^2$ is still subdominant, the growing mode solution to the above equation is
\[\label{eq34}
\sigma\propto a^{\gamma_+}\,\,,\,\,\gamma_+=-\frac34+\frac14\sqrt{9-16c_\sigma}\simeq-2c_\sigma/3\,,
\]
where the last step follows after expanding to first order in $c_\sigma$. As long as \mbox{$m_\sigma^2\simeq c_\sigma H^2$}, the field avoids its oscillatory phase. Therefore, neglecting the kinetic density $\rho_{\rm kin}=\frac12\,\gamma_+^2H^2\sigma^2$ we have $\rho_\sigma/\rho\propto a^{2\gamma_+}$, and hence $\rho_\sigma$ remains always subdominant.

After $H$ decreases enough (for $H\sim m_0$), $\sigma$ begins its oscillatory stage about the origin of its potential, and only when $m_0>H$ the field performs fast oscillations with an amplitude that scales as $\sigma\propto a^{-3/2}$. This scaling continues until the time of reheating, which happens when $H\simeq \Gamma_0$. To secure the survival of the classical $\sigma$ until reheating it suffices to impose that $\Gamma_\sigma<\Gamma(\sigma)$, where $\Gamma_\sigma$ is the decay rate of $\sigma$. Moreover, owing to the smallness of the field oscillations, we can safely neglect any non-perturbative decay of $\sigma$. Note that $\sigma$ oscillates before the time of reheating only if $m_0>\Gamma_0$. Using all the above, the amplitude of the $\sigma$ oscillations at the time of inflaton decay is
\[\label{eq13}
\sigma_{\rm dec}\simeq\sigma_{\rm end}\left(\frac{\Gamma_0}{H_*}\right)^{4c_\sigma/9}\min\left[1,\left(\frac{\Gamma_0}{m_0}\right)^{1-4c_\sigma/9}\right]\,,
\]
where we allow $\Gamma_0/m_0$ to be larger or smaller than one. Using now Eqs.~(\ref{eq12}), (\ref{eq11}) and (\ref{eq13})  we obtain
\[\label{eq19}
\zeta_\sigma\sim\frac{\alpha q r}{2\pi}\,g^{1-q}
\left(\frac{H_*}{M}\right)^q\left(\frac{\Gamma_0}{H_*}\right)^{4qc_\sigma/9}e^{-N_*c_\sigma/3}\min\left[1,\left(\frac{\Gamma_0}{m_0}\right)^{q(1-4c_\sigma/9)}\right]\,,
\]
where we made the replacement $N(k)\to N_*$, for this only introduces a correction of order one when dealing with CMB scales and for $c_\sigma={\cal O}(10^{-1})$.

\subsubsection{Parameter constraints}\label{sec4a}
To further constrain the model parameters we impose the condition $\sigma_{\rm dec}<M$. Using Eq.~(\ref{eq13}) we have
\[
\frac{H_*}{M}<g\left(\frac{H_*}{\Gamma_0}\right)^{4c_\sigma/9}\max\left[1,\left(\frac{m_0}{\Gamma_0}\right)^{1-4c_\sigma/9}\right]\,.
\]
Substituting this into Eq.~(\ref{eq19}) and operating we find
\[\label{eq23}
g>\frac{2\pi\zeta_\sigma}{\alpha q r}\exp(N_*c_\sigma/3)\,,
\]
which is stronger than Eq.~(\ref{eq17}) when $q,r={\cal O}(1)$, $\zeta_\sigma\sim10^{-5}$ and $N_{\rm sr}^p\lesssim50$. Therefore, using $g\leq{\cal O}(1)$, the existence of parameter space for $g$ demands that 
\[
\frac{2\pi\zeta_\sigma}{\alpha qr}\,\exp(N_*c_\sigma/3)<1\,,
\]
from which we derive the allowed range for $c_\sigma$
\[\label{eq21}
0\leq c_\sigma<\frac3{N_*}\,\log\frac{\alpha q r}{2\pi\zeta_\sigma}\,.
\]

\begin{figure}[htbp]
\centering\epsfig{file=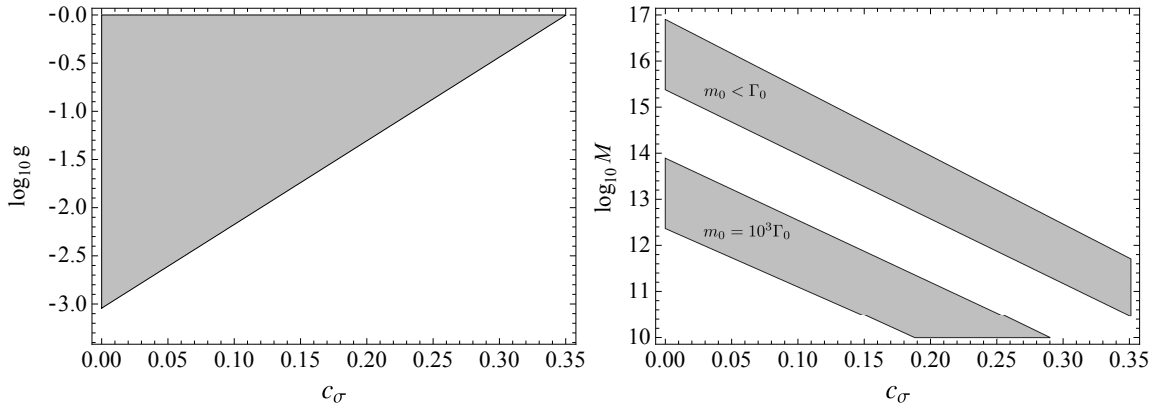,width=15cm}\caption{Range of allowed values for $g$ and $c_\sigma$ (left-hand panel) and for $M$ (righthand panel) after imposing the constraints in Eqs.~(\ref{eq23}) and (\ref{eq21}).}\label{fig4}
\end{figure}

In the left-hand panel of Fig.~\ref{fig4} we depict the parameter range allowed by Eqs.~(\ref{eq23}) and (\ref{eq21}). To build the plot we take $q=2$, $r=1$, $N_*=60$ and $\zeta_\sigma=4.8\times10^{-5}$. After constraining $g$ and $c_\sigma$, we obtain the allowed range for $M$ compatible with Eq.~(\ref{eq19}). Since $M\propto g^{-1/2}$, the minimum [maximum] allowed $M$ corresponds to the maximum [minimum] $g$. The allowed range of $M$ is also determined by $\Gamma_0$, with $M\propto \Gamma_0^{4c_\sigma/9}$ if $\sigma$ does not oscillate before reheating and with $M\propto \Gamma_0$ in the opposite case. To express the range of $M$ in terms of the reheating temperature $T_{\rm rh}$ we assume the sudden decay approximation. The temperature of the radiation bath, formed by the decay products of the inflaton, is then given by $\rho_R=\left(\pi^2g_*/30\right)T_{\rm rh}^4$, where $g_*$ is the effective number of light degrees of freedom. Since $M$ is proportional to positive power of $\Gamma_0$, its minimum [maximum] allowed value corresponds to the minimum [maximum] allowed reheating temperature. For illustration purposes, we consider the range $10^5 {\rm GeV}\leq T_{\rm rh}\leq10^9{\rm GeV}$, which is compatible with current bounds on gravitino overproduction \cite{Kawasaki:2004qu,Kawasaki:2004yh}. The allowed range of $M$ is plotted in the righthand panel of Fig.~\ref{fig4}, where we use $H_*=3\times10^{-5}m_P$.\footnote{Note that stronger bounds for $M$ have been recently shown to arise in modulated preheating \cite{Mazumdar:2015xka}.}  In the plot we have included two cases: $m_0<\Gamma_0$ and $m_0>\Gamma_0$, as indicated. From the above discussion, it follows that the allowed range of $M$ in the second case becomes lowered by a factor $(m_0/\Gamma_0)^{1-4c_\sigma/9}\simeq(m_0/\Gamma_0)$ with respect to the first case.

Our results demonstrate the existence of allowed space for $M$, which affords us to conclude the feasibility of the localized inhomogeneous reheating to account for anomalous hot spots through enhanced underdensities. Moreover, after replacing $M\to\sqrt{qr}\lambda^{-1}m_\phi$, our results also demonstrate the feasibility of our mechanism to account for the Cold Spot through an enhanced overdensity in the last scattering surface. In spite of these encouraging results, we remark that to fully demonstrate that the mechanism explains the Cold Spot it is necessary to 
obtain and compare the predicted temperature profile with the observed one. In particular, the mechanism must reproduce the surrounding hot spot \cite{Zhang:2009qg}. This analysis constitutes the subject of future research.

\subsection{Power deficit at low $\ell$}\label{sec3}
The lack of power in the low multiples of the CMB, currently regarded as one of the most robust anomalies, was first observed by WMAP \cite{Bennett:2003bz,Spergel:2003cb} and later confirmed by Planck \cite{Ade:2013nlj,Ade:2015hxq}. Some of the best known alternatives to account for the power deficit are open inflation \cite{Linde:1999wv,White:2014aua} or, more recently, the generation of an anti-correlated isocurvature perturbation \cite{Contaldi:2014zua,Kawasaki:2014lqa,Kawasaki:2014fwa,Harigaya:2014bsa,Bastero-Gil:2014oga}. But arguably, the simplest and more intuitive alternative to account for the power deficit is to postulate the existence of a phase of fast-roll inflation\footnote{In slow-roll inflation, the spectrum of the curvature perturbation is ${\cal P}_\zeta(k)\propto \epsilon_{\rm sr}^{-1}$, where $\epsilon_{\rm sr}$ is the first slow-roll parameter \cite{Lyth:2009zz}. Therefore, a suppression of power in the largest scales may be accounted for by the corresponding growth in $\epsilon_{\rm sr}$, thus entailing a faster evolution of the inflaton.} \cite{Linde:2001ae,Contaldi:2003zv,Schwarz:2009sj,Boyanovsky:2009xh,Ramirez:2011kk,Lello:2013mfa}. Nevertheless, an important drawback of these models is that the fast-roll stage must finish at about the time when the observable Universe exits the horizon. Although fast-roll inflation can be easily motivated from various particle physics models, the requirement that it finishes just at the right time constitutes something for which there seems to be no compelling reason.

Our scenario for the generation of out-of-equilibrium patches, while requiring a sustained fast-roll stage to generate the initial condition, resembles considerably the essence of the aforementioned models. However, in contrast to them, our framework does not require that the fast-roll stage finishes when the largest observable scales are exiting the horizon. In fact, we consider the case when the fast-roll finishes many $e$-foldings before the observable Universe exits the horizon. In that case, the inflaton perturbation spectrum on CMB scales is the one predicted by slow-roll inflation. Therefore, the power deficit owes exclusively to the isocurvature field $\sigma$, whose initial condition is generated during the epoch of fast-roll.

A feasible alternative to produce a power deficit is to consider an anti-correlated isocurvature perturbation\footnote{Anti-correlated isocurvature perturbations were recently considered in order to alleviate 	the tension between the Planck and BICEP2 data \cite{Contaldi:2014zua,Kawasaki:2014lqa,Kawasaki:2014fwa}, although such tension no longer exists after the results from the joint collaboration Planck/BICEP2 \cite{Ade:2015tva}.}. If in addition to the curvature perturbation imprinted by the inflaton field we consider a matter isocurvature perturbation $S_m$, then temperature fluctuations on large scales are approximated by \cite{Contaldi:2014zua}
\[\label{eq32}
\Big\langle\left(\frac{\Delta T}{T}\right)^2\Big\rangle=\frac1{25}\left({\cal P}_\zeta+4{\cal P}_{S_m}+4{\cal P}_{\zeta S_m}+\frac56\,{\cal P}_t\right)\,,
\]
where ${\cal P}_\zeta$, ${\cal P}_{S_m}$, ${\cal P}_{\zeta S_m}$ and ${\cal P}_t$ are the power spectra of the curvature, isocurvature, cross-correlation and tensor perturbations, respectively. From the above, a matter isocurvature satisfying ${\cal P}_{S_m}+{\cal P}_{\zeta S_m}<0$ reduces the amplitude of temperature fluctuations relative to the adiabatic case, $S_m=0$. We remark, however, that the introduction of a fully anticorrelated matter isocurvature perturbation to account for the power deficit becomes disfavoured after a Bayesian model comparison \cite{Giannantonio:2014rva}.

\subsubsection{Local anticorrelation from a right-handed sneutrino}
According to recent findings, a right-handed sneutrino can play the role of the curvaton and give rise to an anti-correlated CDM/baryon isocurvature, thus suppressing temperature fluctuations on large scales. The necessary condition for this to happen is that curvaton field is \cite{Harigaya:2014bsa}
\[\label{eq84}
\sigma_*={\cal O}(10^{-2})m_P
\]
at the time of horizon crossing. Using this result as a basis, in the following we investigate if our prototype isocurvature field $\sigma$ can satisfy the above requirement while leading to the formation of out-of-equilibrium patches at the end of inflation.

To examine the feasibility of this idea, first we need to consider a fast-roll phase able to generate an initial value $\sigma_{\rm sr}$ sufficiently large so that \mbox{$\sigma_*={\cal O}(10^{-2})m_P$} at the time of horizon crossing for cosmological scales. After that, we must enforce the generation of out-of-equilibrium patches at the end of inflation. As pointed out in Sec.~\ref{sec5}, the emergence of patches greatly depends on $\sigma_{\rm sr}$. Therefore, in principle, nothing guarantees that the appropriate value \mbox{$\sigma_*={\cal O}(10^{-2})m_P$} will also entail the appearance of out-of-equilibrium patches. Below we address the compatibility of these two requirements in detail.

In the first place, we determine the parametric region where out-of-equilibrium patches arise at the end of inflation in the appropriate range of scales. As already explained, to generate the initial value $\sigma_{\rm sr}$ we consider a fast-roll stage, during which the variance $\Sigma^2$ undergoes an unstable growth, and then set $\sigma_{\rm sr}=\Sigma$ at the end of it. As for the subsequent phase of slow-roll, we recall that we consider a primary phase lasting for $N_{\rm sr}^p$ $e$-foldings, after which the largest cosmological scales exit the horizon $N_*$ $e$-foldings before the end of inflation, namely
\[
N_{\rm sr}=N_{\rm sr}^p+N_*\,.
\]
For the purpose of illustration, we study the emergence of patches in the multipole range \mbox{$2\leq\ell\leq40$}, which encompasses the region featuring the power deficit. Since out-of-equilibrium patches are supposed to emerge in sparse regions of the observable Universe, for definiteness, we set their abundance from 1 to 10\% of the observable Universe, namely 
\[\label{eq50}
0.01\leq{\cal F}_{\rm ext}(k_{40})\leq0.1\,.
\]

\begin{figure}[htbp]
\centering\epsfig{file=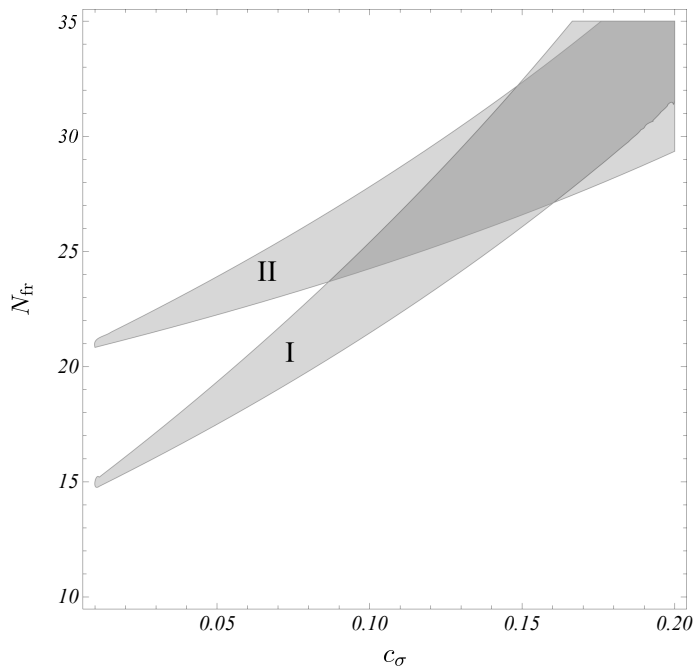,width=9cm}
\caption{Plot of the parametric regions satisfying Eq.~(\ref{eq50}) (region I) and Eq.~(\ref{eq63}) (region II). To build the plot we take $\epsilon=0.3$ to characterize the fast-roll stage, $g=5\times10^{-2}$ for the coupling between $\sigma$ and $\chi$ and $\tau_t=2H_*^{-1}$. The length of the primary phase is kept in the interval $10\leq N_{\rm sr}^p\leq50$.}\label{fig16}
\end{figure}

In Fig.~\ref{fig16}, we plot the parametric region satisfying Eq.~(\ref{eq50}) (region I) while keeping the length of the primary phase in the interval \mbox{$10\leq N_{\rm sr}^p\leq50$}. To build the plot we take a fast-roll stage characterized by $\epsilon=0.3$, a transition time $\tau_t=2H_*^{-1}$, $g=5\times10^{-2}$ and $N_*=60$. As expected, our plot confirms that out-of-equilibrium patches can indeed appear at the end of inflation without having to arrange the end of the fast-roll stage at the time of horizon crossing for the largest observable scales. Next, we take into account the condition in Eq.~(\ref{eq84}). Using $\sigma_{\rm sr}=\Sigma$ and writing $\sigma_*=\sigma_{\rm sr}\exp(-c_\sigma N_{\rm sr}^p/3)$, the condition to generate $\sigma_*$ for a given $N_{\rm sr}^p$ reads $\Sigma=\sigma_*\,\exp(c_\sigma N_{\rm sr}^p/3)$, where $\Sigma$ is given by Eq.~(\ref{eq10}). Since we wish to keep $N_{\rm sr}^p$ within its interval, we constrain the length of the fast-roll stage imposing
\[\label{eq63}
\sigma_*\,\exp\left[\frac{c_\sigma}3\,\min(N_{\rm sr}^p)\right]\leq \Sigma\leq\sigma_*\,\exp\left[\frac{c_\sigma}3\,\max(N_{\rm sr}^p)\right]\,.
\]
The parametric regions satisfying this constraint is depicted in Fig.~\ref{fig16} (region II).

Our results demonstrate that the emergence of out-of-equilibrium patches, with the abundances in Eq.~(\ref{eq50}), is indeed compatible with the condition in Eq.~(\ref{eq84}), necessary for $\sigma$ to play the role of a curvaton and generate a CDM/baryon isocurvature perturbation suppressing temperature fluctuations on large scales. Although we are mainly interested in the emergence of patches in sparse regions of the Universe, we can apply our computation to check its compatibility when out-of-equilibrium patches cover most of the observable Universe, thus giving rise to a statistically homogeneous fluctuation. Therefore, we have checked that there exists plenty parameter space satisfying Eq.~(\ref{eq84}) and $0.95\leq{\cal F}_{\rm ext}(k)\leq1$. In fact, the allowed space is roughly the same as the one displayed in Fig.~\ref{fig16}.

Given our ansatz $\sigma_{\rm sr}=\Sigma$, once we fix $N_{\rm fr}$ and $\epsilon$, the initial value $\sigma_{\rm sr}$ depends on $H_*$, and hence so does $\sigma_*$. According to Planck data, the current bound on the tensor-to-scalar ratio translates into the bound $H_*<3\times10^{-5}m_P$. We thus build Fig.~\ref{fig16} using \mbox{$H_*=10^{-5}m_P$}. Now, if we take a smaller $H_*$, to keep $\Sigma$ fixed (in order to secure that Eq.~(\ref{eq63}) is still satisfied) we require either a larger $N_{\rm fr}$ or a slightly larger $\epsilon$ to compensate. Therefore, the contour defined by Eq.~(\ref{eq63}) in Fig.~\ref{fig16} becomes displaced to larger $N_{\rm fr}$. On the other hand, as long as we choose $\sigma_{\rm sr}=\Sigma$, the fraction ${\cal F}_{\rm ext}(k)$ does not depend on $H_*$, and hence the contour defined by Eq.~(\ref{eq50}) remains unchanged after taking a smaller $H_*$. Consequently, choosing a larger $N_{\rm fr}$ to compensate for a smaller $H_*$ results in a reduction of the space satisfying Eqs.~(\ref{eq50}) and (\ref{eq63}). We have checked that, within the range of $N_{\rm fr}$ and $c_\sigma$ shown, these constraints become incompatible for $H_*\simeq 10^{-6}m_P$. Although this gives a narrow margin for $H_*$, we remark that this result is derived for particular values the parameters $g$ and $\epsilon$. To make the bounds in Eqs.~(\ref{eq50}) and (\ref{eq63}) compatible again for $H_*<10^{-6}m_P$, we need to push the contour defined by Eq.~(\ref{eq50}) to higher $N_{\rm fr}$. This can be achieved by choosing a smaller coupling $g$. Since this takes $\sigma_c\propto g^{-1}$ to larger field values, a larger initial value $\sigma_{\rm sr}$, and hence a larger $N_{\rm fr}$, is required to satisfy Eq.~(\ref{eq50}). On the other hand, the constraint in Eq.~(\ref{eq63}) is independent of $g$, as field interactions play no role in the generation of the initial condition $\sigma_{\rm sr}$, or equivalently $\sigma_*$.  Consequently, a smaller $g$ affords us to comply with the constraints in Eqs.~(\ref{eq50}) and (\ref{eq63}) while using a smaller $H_*$. We have checked that in order to find substantial allowed space with $H_*=10^{-6}m_P$ it suffices to take $g<5\times10^{-2}$. It is worth stressing that, as explained in Sec.~\ref{sec11}, the probability for the emergence of patches in the range of interest $c_\sigma={\cal O}(10^{-1})$ is rather insensitive to the magnitude of the coupling $g$ as long as $H_*\tau_t\geq{\cal O}(1)$. We thus conclude that the mechanism here described can accommodate a smaller $H_*$ without harming the naturalness of the emergence of patches at the end of inflation.

Finally, in order to complete the model, we must secure that the candidate curvaton field has the necessary couplings to the $\chi$-like degrees of freedom in Eq.~(\ref{eq27}). But such couplings are already present, for example, in minimal hybrid-like models \cite{Antusch:2004hd}. We thus conclude that, in principle, a right-handed sneutrino with a hybrid-like potential can become a successful curvaton field imprinting an anti-correlated isocurvature perturbation in sparse regions of the Universe only.

\subsection{The breaking of statistical isotropy}
Over the last decade, a number of observations have questioned the long-standing assumption of statistical isotropy of the CMB. Notable examples of this are the alignment between the preferred axis of the quadrupole and octopole, an observation usually referred to as the axis of evil \cite{deOliveira-Costa:2003utu,Schwarz:2004gk,Land:2006bn}, and the presence of a hemispherical or dipole modulation \cite{Eriksen:2003db,Hansen:2004vq,Eriksen:2007pc,Hoftuft:2009rq,Akrami:2014eta}. However, it is still not clear whether such observations originate from a preferred direction in the Universe \cite{Ade:2013nlj,Ade:2015hxq,Schwarz:2015cma}. Although several models have been explored to explain these observations while resorting to scalars \cite{Erickcek:2009at,Lyth:2013vha,Mazumdar:2013yta,Kanno:2013ohv,McDonald:2013qca,McDonald:2014lea}, cosmological vector fields are natural candidates to account for such obserations since they can single out a preferred direction in space. Therefore, in this section we consider the intervention of a vector field to break the statistical isotropy of the CMB \cite{Dimopoulos:2006ms,Dimopoulos:2008yv,Kanno:2008gn,Dimopoulos:2009am,Dimopoulos:2009vu,Gumrukcuoglu:2010yc,Watanabe:2010fh,Dimopoulos:2011ws,Murata:2011wv,Soda:2012zm,Lyth:2012vn,Bartolo:2012sd,Chen:2013eaa}. The risk in this case, however, is that the vector field results in an anisotropic expansion in excess of the current observational bounds. To quantify the anisotropy, it is usual to parametrize the spectrum of the curvature perturbation $\zeta$ as \cite{Ackerman07}
\[\label{eq68}
{\cal P}_\zeta(\mbox{\boldmath $k$})={\cal P}_\zeta^{\rm iso}(k)
[1+g_*(\mbox{\boldmath $d\cdot\hat k$})^2]\,,
\]
where ${\cal P}_\zeta^{\rm iso}(k)$ denotes the isotropic part of the power spectrum, \mbox{\boldmath $d$} is the unit vector signaling the preferred direction, $\mbox{\boldmath $\hat k$}\equiv\mbox{\boldmath $k$}/k$ is the unit vector along the wavevector \mbox{\boldmath $k$} and $g(k)$ is the anisotropy parameter. The analysis of the data from the WMAP and Planck satellites results in the constraint \cite{Hanson:2010gu,Ramazanov:2013wea,Kim:2013gka,Ade:2015hxq,Ade:2015lrj}
\[\label{eq16}
g_*\lesssim2\times10^{-2}\,,
\]
which represents a very strong restriction on the contribution of vector fields to the primordial perturbation spectrum.

It is convenient to stress, however, that the bound in Eq.~(\ref{eq16}) is obtained under the assumption of spatial homogeneity of the vector perturbation, and hence it cannot be applied in a straightforward manner if the perturbation is very inhomogeneous. Bearing this caveat in mind, our goal for this section is to explore a mechanism to generate such an inhomogeneous perturbation. To do so, we study the emergence of out-of-equilibrium patches in a cosmological vector field and then investigate if this can generate an observable direction-dependent contribution to $\zeta$ in isolated patches of the CMB.

\subsubsection{A toy model for local vector perturbations}
In order to keep our model in the simplest, we consider the well-studied case of a massive vector field $A_\mu$ with a varying kinetic function \cite{Dimopoulos:2009am}
\[\label{eq66}
{\cal L}_A=-\frac14\,f\,F_{\mu\nu}F^{\mu\nu}+\frac12\,m^2A_\mu A^\mu\,.
\]
Arguably, this is the simplest stable theory in which massive vector fields can be produced during inflation \cite{Himmetoglu:2008zp,Himmetoglu:2008hx,Dimopoulos:2009vu}. In order for the vector field to be substantially produced during inflation, the kinetic function of the vector field and its mass are allowed to have a time-dependence parametrized by 
\[\label{eq53}
f\propto a^\alpha\quad\textrm{and}\quad m\propto a^\beta\,.
\]
Using this model, a successful vector curvaton mechanism can be built, with a scale-invariant spectrum of vector perturbations for appropriate values of $\alpha$ and $\beta$ (see \cite{Dimopoulos:2011ws} for a review). 

According to the discussion in \cite{Namba:2012gg}, the kinetic function and mass of the vector should be determined by an additional field, which the author takes to be the inflaton. For our purposes, however, it suffices to consider the case when only the kinetic function becomes modulated by an additional field
\[\label{eq77}
f=f(\sigma)\,,
\]
which we take to be our prototype isocurvature field. Therefore, apart from $\sigma$, we also require the additional $\chi$ sector in Eq.~(\ref{eq27}) and the appropriate initial conditions so that $\sigma$ features a distribution of out-of-equilibrium patches at the end of inflation.

At this point it is convenient to emphasize that although deviations from scale-invariance can be found in the perturbation spectrum of $A_\mu$ (depending on both $f(\sigma)$ and the dynamics of $\sigma$), for our purposes such deviations do not constitute a concern. This is because the nearly scale-invariance of the spectrum becomes necessary only if the vector field is to account for most of the primordial spectrum. Since our goal is to construct a model allowing the vector field to imprint its perturbation only in sparse regions of the Universe, imposing the nearly scale-invariance of the spectrum is unnecessarily constraining. With this remark in mind, however, we choose to stick to scale-invariance simply because this affords us to keep our analysis in the simplest.

In the following, rather than using $A_\mu$ we use the physical vector field $\mbox{\boldmath$W$}\equiv\sqrt f\mbox{\boldmath$A$}/a$, taking its spatial component oriented along the z-axis. Using $f\propto a^\alpha$, the evolution equation for the homogeneous component of \mbox{\boldmath$W$} (here denoted by $W$) during slow-roll inflation can be approximated by
\[\label{eq56}
\ddot W+3H\dot W+\left[-\frac14(\alpha+4)(\alpha-2)H^2+M^2\right]=0\,,
\]
where $M\equiv m/\sqrt{f}$ is the mass of the canonically normalized field. We assume $M\ll H$ so that the vector field can be produced during inflation. Furthermore, if we assume equipartition of the energy at the onset of inflation, the evolution of the vector field is well approximated by $W\propto a^{-3}$ for $\alpha\simeq-4$ and $W\simeq{\rm const.}$ for $\alpha\simeq2$, where $\alpha=-4,2$ are the cases corresponding to scale-invariance of the vector perturbation \cite{Dimopoulos:2009am,Dimopoulos:2009vu}. In that case, the energy density of the vector field \mbox{$\rho_A=\rho_{\rm kin}+V_A$}, where the kinetic and potential energy densities are \mbox{$\rho_{\rm kin}=\frac12[\dot W+\left(1-\alpha/2\right)HW]^2$} and $V_A=-\frac12m_A^2A_\mu A^\mu$, remains approximately constant during inflation, with
\[\label{eq51}
\rho_A\simeq M_*^2W_*^2\,.
\]

Regarding the perturbation spectrum, since the vector field is massive we must quantize three degrees of freedom: two transverse and one longitudinal. After defining the transverse left ($L$) and right ($R$) and longitudinal ($\parallel$) polarizations vectors, the perturbation spectrum for each polarization is
\[\label{eq52}
{\cal P}_{L,R}=\left(\frac{H}{2\pi}\right)^2\quad\,,\quad{\cal P}_\parallel=\left(\frac{H}{2\pi}\right)^2\left(\frac{H}{3M}\right)^2\,.
\]
If $M<3H$ by the end of inflation, the vector perturbation is dominated by the longitudinal mode, and hence it becomes highly anisotropic. As pointed out in \cite{Dimopoulos:2009am,Dimopoulos:2009vu,Dimopoulos:2011ws}, the compatibility of observations, i.e. $g_*\lesssim10^{-2}$, with a highly anisotropic perturbation only allows the vector field to give a subdominant contribution to the curvature perturbation. Using now Eqs.~(\ref{eq51}) and (\ref{eq52}) and introducing $\hat m\equiv M_{\rm end}$, we can compute the fractional field perturbation at the end of inflation, obtaining
\[\label{eq55}
\left(\frac{\delta W}{W}\right)_{\rm end}\simeq\frac{3H_*^2}{2\pi(\rho_A)_{\rm end}^{1/2}}\,.
\]

Since the rolling of $\sigma$ during slow-roll inflation is supposed to induce the scaling in Eq.~(\ref{eq53}), the above result is the expected one in patches where $\sigma$ remains in its slow-roll phase until the end of inflation, namely in out-of-equilibrium patches. On the other hand, given that a successful curvaton mechanism can be built for this model \cite{Dimopoulos:2009am,Dimopoulos:2009vu,Dimopoulos:2011ws}, we conclude that the vector field can contribute to the total curvature perturbation imprinted in the CMB in out-of-equilibrium patches. 

We focus now on spatial patches where $\sigma$ reaches its oscillatory regime during inflation. Since $A_\mu$ is coupled to $\sigma$ through the kinetic function, to study the consequences of the transition to the oscillatory regime for $W$ we need to specify a particular form of $f$. Rather general forms of $f(\sigma)$ can be easily motivated from supergravity: $f(\sigma)\propto(\sigma/\tilde M)^{2n}$ \cite{Dimopoulos:2007zb}, or from dilaton electromagnetism in string theory: $f(\sigma)=\exp(\lambda\,\sigma/m_P)$ \cite{Bamba:2003av,Bamba:2004cu}. Although we do not pursue a detailed analysis of any of these models, in the following we use the fact that they give rise to a canonically normalized kinetic term, $f=1$, for sufficiently small $\sigma$. In turn, this is certainly expected to happen once $\sigma$ engages into its oscillatory regime during inflation, for in that case the amplitude of the oscillations about $\sigma=0$ becomes exponentially suppressed. Therefore, the computation below applies to those patches where $\sigma$ has been oscillating for a sufficient number of $e$-foldings until the end of inflation.

Although the lack of a particular model for $f$ naturally bounds the reach of our results, we expect that the way in which the scaling regime for $f$ comes to an end has little impact on the fractional perturbation $(\delta W/W)_{\rm end}$. This is so because on sufficiently superhorizon scales (and this is the case of CMB scales when the interactions of $\sigma$ become important) the perturbation modes of \mbox{\boldmath$W$} approximately obey the same equation as the homogeneous field $W$. Thus, the fractional perturbation $(\delta W/W)_{\rm end}$ is not expected to be significantly different from the one in Eq.~(\ref{eq55}). However, and in contrast to this, the evolution of $\rho_A$ undergoes a critical change. To see this, we write the evolution equation for the homogeneous $W$ using $f=1$. Since the latter implies $\alpha=0$, from Eq.~(\ref{eq56}) we obtain 
\[\label{eq54}
\ddot W+3H\dot W+(2H^2+\hat m^2)W=0\,,
\]
where for simplicity we assume that $M$ at the end of the scaling regime, when reached before the end of inflation, coincides with $M_{\rm end}$. As already explained, we allow for the possibility that $\hat m^2<H^2$ in order to obtain a strongly anisotropic perturbation spectrum in the patches where $\sigma$ remains in its slow-roll phase until the end of inflation. Solving the above with $\hat m\simeq{\rm const.}$ we find
\[
W\propto a^{-3/2}a^{\sqrt{1/4-\hat m^2/H^2}}\simeq a^{-1}\,,
\]
where we neglect the correction from $\hat m^2/H^2$. Using this we obtain 
\[\label{eq47}
\rho_A\simeq\frac12\left(\dot W+HW\right)^2+\frac12\hat m^2W^2\simeq\frac12\left[\hat m^2+H^2\left(\frac{\hat m^2}{H^2}\right)^2\right]W^2\propto a^{-2}\,,
\]
which is to be contrasted with the result in Eq.~(\ref{eq51}), where $\rho_A$ remains constant.

\subsubsection{Implications for the curvature perturbation}
To transfer the modulation in the vector field $A_\mu$ to the curvature perturbation we resort to the vector curvaton paradigm. Following \cite{Dimopoulos:2009am}, the curvature perturbation including the contribution from the vector field is
\[\label{eq42}
\zeta=(1-\hat\Omega_A)\zeta_{\rm rad}+\hat\Omega_A\zeta_A\,,
\]
where $\zeta_{\rm rad}$ is the curvature perturbation present in the radiation dominated universe after inflation, $\zeta_A$ is the curvature perturbation in the vector field, $\hat\Omega_A\equiv3\Omega_A/(4-\Omega_A)$ sets the relative contribution of each component to the total curvature perturbation and $\Omega_A\equiv\rho_A/\rho$ is the density parameter of the vector field.

After inflation, we take the kinetic function to be normalized to $f=1$. In that case, if the vector field is heavy at the end of inflation ($\hat m>H$), its energy density scales as $\rho_A\propto a^{-3}$. Since $\Omega_A\propto a$ during radiation domination, the vector field can come to dominate, or nearly dominate, the energy density of the Universe. However, if $\hat m<H$, which results in a highly anisotropic spectrum, we have $\rho_A\propto a^{-4}$. Then $\Omega_A$ remains constant during the radiation dominated epoch, starting to scale as $\Omega_A\propto a$ only after $H$ has decreased enough so that $M>H$, when the field begins to oscillate. Since this last case $\hat m<H$ is the one of interest to us, we must secure that the oscillations of $W$ begin early enough so that $\rho_A$ can come to dominate, or nearly dominate, the energy density Universe. In fact, as demonstrated in \cite{Dimopoulos:2009am}, the model parameters can be chosen to allow the vector field to give a small, highly anisotropic contribution to the total curvature perturbation. Therefore, this result affords us to invoke the existence of a successful vector curvaton mechanism in out-of-equilibrium patches.

After the background radiation is sufficiently redshifted, the vector field imprints its curvature perturbation when it decays into radiation. At this time (labeled by ``dec''), the curvature perturbation in the vector field can be written as
\[\label{eq82}
\zeta_A=\frac13\left(\frac{\delta\rho_A}{\rho_A}\right)_{\rm dec}\simeq\frac23\left(\frac{\delta W}{W}\right)_{\rm end}\,.
\]
Using now Eqs.~(\ref{eq55}) and (\ref{eq42}) and taking $\Omega_A\lesssim1$ at the time of decay, we can approximate the anisotropic part of the curvature perturbation as follows
\[\label{eq45}
\zeta_{\rm ani}(\mbox{\boldmath$x$})=\hat\Omega_A\zeta_A\simeq\frac{\sqrt3}{4\pi}\left[\frac{(\Omega_A)_{\rm dec}}{(\Omega_A)_{\rm end}}\,\frac{H_*}{m_P}\right]\Omega_{A,{\rm end}}^{1/2}(\mbox{\boldmath$x$})\,.
\]
If the scaling in Eq.~(\ref{eq53}) finishes in out-of-equilibrium patches right after the end of inflation, then $\rho_A$ decays everywhere at the same rate from the end of inflation onwards. As a result, the factor in brackets in Eq.~(\ref{eq45}) is independent of the spatial location. However, since $(\rho_A)_{\rm end}$ does depend on the spatial location [c.f. Eqs.~(\ref{eq51}) and (\ref{eq47})], the anisotropic part $\zeta_{\rm ani}$ becomes spatially modulated accordingly. Therefore, using a superscript ${\rm (out)}$ to denote out-of-equilibrium patches and ${\rm (osc)}$ for patches where $\sigma$ oscillates before the end of inflation, we can write
\[\label{eq78}
\zeta^{(\rm osc)}_{\rm ani}(\mbox{\boldmath$x$})=\left(\frac{\rho_A^{(\rm osc)}}{\rho_A^{(\rm out)}}\right)_{\rm end}^{1/2}\zeta^{(\rm out)}_{\rm ani}\,.
\]

It is important to emphasize that $\zeta^{(\rm out)}_{\rm ani }$ has the same value in all out-of-equilibrium patches, whereas $\zeta^{(\rm osc)}_{\rm ani}$ depends on the spatial location \mbox{\boldmath$x$}. Indeed, its magnitude in a particular patch depends on the time when the transition to the oscillatory phase happens. Since this time is a stochastic variable, 
so it is $\zeta^{(\rm osc)}_{\rm ani}$, which then has a probability density associated to it. In order to estimate the ratio $\rho_A^{({\rm osc})}/\rho_A^{({\rm out})}$ we denote by $\hat a(\mbox{\boldmath $x$})$ the scale factor when the scaling in Eq.~(\ref{eq53}) finishes at $\mbox{\boldmath$x$}$. Then, from Eq.~(\ref{eq47}) we find $(\rho_A)^{({\rm osc})}_{\rm end}\sim\exp[-2N_{\rm osc}(\mbox{\boldmath$x$})]\,(\rho_A)^{({\rm out})}_{\rm end}$, where $N_{\rm osc}(\mbox{\boldmath$x$})$ is a stochastic variable giving the remaining number of $e$-foldings when the scaling regime finishes at \mbox{\boldmath$x$}. Using this along with Eq.~(\ref{eq45}) we obtain
\[\label{eq67}
\zeta^{({\rm osc})}_{\rm ani}(\mbox{\boldmath$x$})\sim\exp[-N_{\rm osc}(\mbox{\boldmath$x$})]\,\zeta^{({\rm out})}_{\rm ani}\,.
\] 

Finally, appealing now to the aforementioned existence of a successful curvaton mechanism in out-of-equilibrium patches, we can always find values for the model parameters so that the contributed curvature perturbation becomes observable, i.e. $\zeta^{({\rm out})}_{\rm ani}\sim10^{-5}$, in which case Eq.~(\ref{eq67}) implies that $\zeta^{({\rm osc})}_{\rm ani}(\mbox{\boldmath$x$})$  is too small to be observable.

\subsubsection{The prospect of a vector spot}
As previously explained, the constraint on the anisotropy parameter in Eq.~(\ref{eq16}) does not apply if the vector field perturbation, and hence the corresponding curvature perturbation $\zeta_{\rm ani}(\mbox{\boldmath$x$})$, is statistically inhomogeneous. Our relation in Eq.~(\ref{eq67}) demonstrates that this is precisely the case for the model under study. As a result, we can envisage now a situation in which $\zeta_{\rm ani}(\mbox{\boldmath$x$})$ becomes observable only in sparse regions of the Universe. Since the contribution $\zeta_{\rm ani}^{\rm (osc)}(\mbox{\boldmath$x$})$ is too small to be observable in that case, the corresponding prediction for the anisotropy parameter $g_*^{(\rm osc)}(\mbox{\boldmath$x$})$ may easily respect Eq.~(\ref{eq16}). Moreover, when the vector perturbation is highly anisotropic, which happens for $\hat m<H$ in out-of-equilibrium patches, the corresponding prediction for $g_*^{(\rm out)}$ can become much larger than $g_*^{(\rm dec)}(\mbox{\boldmath$x$})$. Now, since we are focusing on the anisotropic part of $\zeta$, using Eqs.~(\ref{eq68}) and (\ref{eq67}) we obtain the relation
\[\label{eq62}
g_*^{({\rm osc})}(\mbox{\boldmath$x$})\sim\exp[-2N_{\rm dec}(\mbox{\boldmath$x$})]g_*^{({\rm out})}\,.
\]

In principle, this relation can be put forward to motivate the search for \emph{isolated regions} in the Universe (in particular in the last scattering surface) where $g_*$ is in excess of the observational bound $g_*\lesssim2\times10^{-2}$. This affords us to hypothesize the existence of a \emph{Vector Spot} in the CMB. In this sense, it is worth recalling that although non-Gaussianity of the CMB is strongly constrained by observations, with $|f_{NL}|\lesssim{\cal O}(1)$ \cite{Ade:2015ava,Ade:2015lrj}, such a straitjacket does not preclude the emergence of the large non-Gaussian fluctuation known as the Cold Spot. Simili modo, one might regard the observational bound on $g_*$ as analogous to those on $f_{\rm NL}$, thus allowing for local violations of the former, as suggested by Eq.~(\ref{eq62}). Needless to say that, owing to the local nature of the fluctuation here discussed, an eventual detection of a Vector Spot would be inescapably affected by the same \emph{a posteriori} issues as the Cold Spot.

\section{Conclusions}
In this paper, we have investigated a framework to provide a common origin for the large-angle anomalies in the Cosmic Microwave Background. This is based on the generation of statistical inhomogeneous fluctuations in isocurvature fields of mass $m\sim H$, which are then interpreted as the seeds for the large-angle anomalies. Given that some anomalies can have an origin different than others, our framework envisages that they are realized through different mechanisms using different isocurvature fields, and hence our framework should be able to generate statistical inhomogeneous fluctuations in a number of fields. To secure the abundance of candidate fields, we focus on scalar fields with masses $m\sim H$, since these are generic in supergravity theories. In particular. The paper is then divided into two parts, the generation of statistical inhomogeneity in a single isocurvature field and, secondly, the study of different mechanisms to provide a realization for different anomalies.

To address the first part, we investigate the dynamics and observational implications of a single isocurvature field, $\sigma$, when its initial conditions are generated during a sustained stage of fast-roll inflation and focusing on $c_\sigma={\cal O}(10^{-1})$. We emphasize that the fast-roll stage utilized in our framework bears an important difference with respect to similar stages considered in the literature. In our case, the fast-roll stage gives way to slow-roll inflation many $e$-foldings before the largest cosmological scales exit the horizon, and hence the curvature perturbation imprinted by the inflaton on CMB scales is the one predicted in slow-roll inflation. We show that if $\epsilon$ is sufficiently large during the fast-roll so that \mbox{$3-2\nu<0$} (see Eq.~(\ref{eq10})), the field variance $\Sigma^2$ undergoes an unstable growth. If this persists for long enough, at the onset of the slow-roll the magnitude of a typical fluctuation can be much larger than the amplitude of the equilibrium fluctuations in de Sitter space. Consequently, a large $\epsilon$ during the fast-roll stage induces an out-of-equilibrium configuration in the classical field at the onset of the slow-roll. In Fig.~\ref{fig3b} we  demonstrate this behavior for fast-roll stages of moderate length, namely $N_{\rm nsr}=20-30$ $e$-foldings, and for $\epsilon=0.2-0.3$ and $c_\sigma=0.15$.
 
To describe the dynamics of $\sigma$ during the subsequent slow-roll stage, which we allow to happen with cosmological scales still within the horizon, we consider the interaction of $\sigma$ with other scalar degrees of freedom, $\chi$, through a term of the form $g^2\sigma^2\chi^2$. In that case, the evolution of $\sigma$ greatly depends on the initial value $\sigma_{\rm sr}$ at the onset of the slow-roll. If $\sigma_{\rm sr}>\sigma_c$, the $\chi$ field becomes heavy and contributes to the effective mass of $\sigma$ only through quantum corrections, which we disregard. Owing to the fluctuations undergone by $\sigma$ during slow-roll, its interactions with the $\chi$ field become dynamically important at different times in different locations. As a result, it becomes possible to find regions where $\sigma$ remains in its slow-roll stage until the end of inflation, whereas in other regions $\sigma$ is already oscillating before the end of inflation. Due to the different scaling undergone by $\sigma$, Eqs.~(\ref{eq20}) and (\ref{eq22}), at the end of inflation its value greatly differs from one class of regions to the other. From the necessary condition for this behavior to appear, i.e. $\sigma_*>\sigma_c$, we obtain a constraint for the coupling $g$, Eq.~(\ref{eq17}), and another one for the length of the primary phase $N_{\rm sr}^p$, Eq.~(\ref{eq44}). Restricting ourselves to $c_\sigma={\cal O}(10^{-1})$, we find that primary slow-roll inflation can only last for a few tens of $e$-foldings, at most. This is an important result, for it shows that the emergence of out-of-equilibrium patches entails a significant constraint on the length of primary inflation. 
 
To estimate the stochastic properties of the distribution of out-of-equilibrium patches, Eqs.~(\ref{eq3}), (\ref{eq4}) and (\ref{eq6}), we utilize the absorbing barrier approximation, implemented through the boundary condition in Eq.~(\ref{eq26}). This affords us to use a simple analytical expression to obtain a rough estimate on the abundance of out-of-equilibrium patches. Also, to estimate the scale-dependent behavior of the distribution of patches, we simplify their geometry approximating it by a sphere. We then find that the relative number density of out-of-equilibrium patches decreases very rapidly as we decrease the scale, Eqs.~(\ref{eq25b}) and (\ref{eq61}) (see also Figs.~\ref{fig2} and \ref{fig9abc}). This suggests that if out-of-equilibrium patches have observable consequences on scales corresponding to CMB anomalies, one can expect that the significance of any anomalous signature (caused by the existence of patches) on smaller scales will quickly reduce.

Despite its usefulness, the absorbing barrier approximation involves a questionable assumption, for it entails the instantaneous transition of $\sigma$ to its oscillatory regime. As a result, this approximation underestimates the abundance of out-of-equilibrium patches and, more importantly, results in a \emph{fictitious} magnification of the level of tuning necessary for the emergence of out-of-equilibrium patches, Eqs.~(\ref{eq33}) and (\ref{eq46}) (see Fig.~\ref{fig13}). Then, we explore a phenomenological model, Eqs.~(\ref{eq76}) and (\ref{eq70}) (see Fig.~\ref{fig14}), in which the transition to the oscillatory phase of $\sigma$ takes place in the finite timescale $\tau_t$. We set $H_*\tau_t={\cal O}(1)$ and reevaluate the abundance of patches at the end of inflation, Eq.~(\ref{eq64}) (see Fig.~\ref{fig14}), and the level of tuning, confirming that it becomes significantly alleviated, Eqs.~(\ref{eq33}) and (\ref{eq41}) (see Fig.~\ref{fig15}). Our results show that if the appropriate conditions are given, i.e. a fast-roll stage with the appropriate $N_{\rm fr}$ and $\epsilon$, the probability for the emergence out-of-equilibrium patches becomes independent of $g$ in the range shown in Eq.~(\ref{eq79}), thus encompassing our range of interest $c_\sigma\geq{\cal O}(10^{-1})$ (see Fig.~\ref{fig15}). More importantly, this probability remains at the percent level, which then offers an avenue to explain CMB anomalies without having to resort to alternative hypotheses more unlikely that the very existence of anomalies. 

In the second part of the paper, we explore the observational implications that the emergence of out-of-equilibrium patches may have in relation to some of the CMB anomalies. Regarding the Cold Spot we develop a local version of the inhomogeneous reheating mechanism. The essence of the mechanism is that out-of-equilibrium patches (where $\sigma$ retains a relatively large value $\sigma\sim\sigma_c$) are the only regions where the inflaton's decay rate becomes sufficiently perturbed to affect CMB temperature fluctuations. Depending on the monotonic behavior of $\Gamma(\sigma)$, we can obtain enhanced matter underdensities (as in Eq.~(\ref{eq2b})), giving rise to anomalous hot spots when they appear in the last scattering surface. On the other hand, we can also obtain enhanced matter overdensities (as in Eq.~(\ref{eq14})), which results in anomalous cold spots when they appear in last scattering surface. We compute the contribution to the curvature perturbation $\zeta_\sigma$ in Eq.~(\ref{eq19}) using the decay rate in Eq.~(\ref{eq2b}), assuming a matter dominated Universe until reheating and allowing for the possibility of oscillations of $\sigma$ before reheating. Moreover, we allow the reheating temperature to vary in the interval $10^5{\rm GeV}\leq T_{\rm rh}\leq10^9{\rm GeV}$, where gravitino production is less problematic. The existence of allowed space for $M$ (regardless of the existence of an oscillatory phase for $\sigma$ before reheating), found after imposing that $\zeta_\sigma$ becomes observable and restricting $g$ and $c_\sigma$ to their allowed ranges, Eqs.~(\ref{eq23}) and ({\ref{eq21}), demonstrates the feasibility of the local inhomogeneous reheating to account for the Cold Spot (see Fig.~\ref{fig4}) by means of an enhanced overdensity in the last scattering surface.

To account for the power deficit at low multipoles, we seek guidance from recent developments and investigate whether our prototypic isocurvature field $\sigma$ can play the role of a curvaton field generating an anti-correlated CDM/baryon isocurvature perturbation.  Our results, summarized in Fig.~\ref{fig16}, demonstrate that out-of-equilibrium patches (with angular sizes corresponding to $2\leq\ell\leq40$ and the abundances in Eq.~(\ref{eq50})) can indeed arise even after a non-negligible phase of primary slow-roll inflation following the fast-roll stage. In particular, we illustrate our results allowing $N_{\rm sr}^p$ to take on values in the interval $10\leq N_{\rm sr}^p\leq 50$. We find ample space in the range of interest $c_\sigma={\cal O}(10^{-1})$ roughly corresponding to $15\lesssim N_{\rm fr}\lesssim30$ and $\epsilon=0.3$  (see region I in Fig.~\ref{fig16}). Also, we depict the allowed parameter space after imposing Eq.~(\ref{eq84}), necessary so that a right-handed sneutrino curvaton results in an anti-correlated CDM/baryon isocurvature perturbation. Allowing $N_{\rm sr}^p$ to vary in the interval $10\leq N_{\rm sr}^p\leq 50$, the condition in Eq.~(\ref{eq84}) translates into Eq.~(\ref{eq63}), which determines region II. The existence of allowed space after enforcing Eqs.~(\ref{eq50}) and (\ref{eq63}) demonstrates that, indeed, the emergence of out-of-equilibrium patches of $\sigma$ can account for the power deficit at low $\ell$ by identifying $\sigma$ with a right-handed sneutrino.

As a final application, we consider the breaking of statistical isotropy due to a cosmological vector field $A_\mu$. In spite of its phenomenological interest, a cosmological vector field may give rise to an anisotropic contribution to the curvature perturbation, which is strongly constrained by observations, Eqs.~(\ref{eq68}) and (\ref{eq16}). Nevertheless, such a constraint can be avoided if the vector field imposes an statistically inhomogeneous perturbation. To investigate this idea we aim at a local version of the vector curvaton mechanism, using a stable massive vector field model as a basis, Eqs.~(\ref{eq66}) and (\ref{eq53}), coupled to our prototype field $\sigma$ through the kinetic function $f(\sigma)$ and taking for granted the necessary conditions to secure the emergence of out-of-equilibrium patches in $\sigma$ at the end of inflation. In out-of-equilibrium patches, the evolution of $A_\mu$ is the same as for the successful vector curvaton used as a basis. Consequently, $A_\mu$ can imprint a sizable, highly anisotropic contribution $\zeta_{\rm ani}(\mbox{\boldmath$x$})$ to the curvature perturbation if it remains light until the end of inflation. The evolution of $A_\mu$, on the other hand, becomes very different in patches where $\sigma$ reaches its oscillatory regime before the end of inflation. Identifying the onset of the oscillatory stage with the end of the scaling regime in Eq.~(\ref{eq53}), we show that the energy density $\rho_A$ becomes suppressed after $\sigma$ starts oscillating, Eq.~(\ref{eq47}). The consequence of this suppression is that the corresponding contribution $\zeta_{\rm ani}(\mbox{\boldmath$x$})$ becomes unobservable, Eqs.~(\ref{eq45})-(\ref{eq67}). Therefore, by coupling $A_\mu$ to our prototype field $\sigma$ we discover that it is indeed possible to imprint a highly anisotropic contribution to the curvature perturbation in sparse regions of the Universe. Additionally, as a natural spin-off emerging from this conclusion, we may conjecture the existence of a \emph{Vector Spot} in the Cosmic Microwave Background. Nevertheless, the eventual identification of such a vector spot would be affected by the same a priori issues as the Cold Spot.

\section*{Acknowledgements}
The author wishes to thank M. Bastero-Gil and E. Mart\'inez-Gonz\'alez for comments and discussions. The author is supported by COLCIENCIAS grant No. 110656399958.

\appendix
\section{On boundary conditions}\label{sec2}
The use of boundary conditions different from those in Eq.~(\ref{eq24}) can be motivated as follows. As discussed in Sec.~\ref{sec8}, $\sigma$ becomes suppressed after entering the oscillatory regime during inflation, and hence we expect that it does not have observational consequences. Therefore, we are mainly interested in out-of-equilibrium patches, where $\sigma$ retains a large value until the end of inflation. To obtain the main qualitative features on the statistics of these regions, we make the following simplifying assumption: If field interactions (responsible for the transition to the oscillatory phase) become important at any time during inflation in a given patch, i.e. whenever $\sigma=\sigma_c$, we will assume that by the end of inflation the field has been oscillating for a sufficiently long time so that its typical value becomes negligible in that patch. Such an assumption can be easily conveyed to the probability density $P_k(\sigma,t)$ by imposing the so-called \textit{absorbing barrier} boundary conditions \cite{Chandrasekhar:1943ws,CoxMiller}
\[\label{eq26}
P_k(\sigma_c,t)=0\,.
\]
In the context of inflation, boundary conditions of this sort have been discussed in \cite{Lorenz:2010vf,Sanchez:2012tk}.

An important caveat to have in mind is that Eq.~(\ref{eq26}) implies an instantaneous transition to the oscillatory regime. However, since this transition is caused by the production of superhorizon fluctuations of the $\chi$ field, the natural expectation is that the transition takes place in the Hubble timescale. Consequently, although the absorbing barrier approximation constitutes a convenient computational tool, it will only provide us with an estimate on the statistics of out-of-equilibrium patches. In particular, since Eq.~(\ref{eq26}) causes the disappearance of the classical field earlier than expected, we can anticipate that the absorbing barrier approximation underestimates the abundance of out-of-equilibrium patches. This is confirmed in Sec.~\ref{sec7}, where we seek to overcome this drawback by proposing a phenomenological model whereby the transition of $\sigma$ to the oscillatory phase occurs in the Hubble timescale. In any case, it is worth emphasizing too that the approximation examined in this section provides the minimal case scenario to study the implications of out-of-equilibrium patches.

Yet another issue to take care of when imposing Eq.~(\ref{eq26}) is the validity of the solution in Eq.~(\ref{eq29}), for this was obtained imposing the boundary condition Eq.~(\ref{eq24}). Therefore, since we evaluate Eq.~(\ref{eq29}) at the end of inflation to obtain Eq.~(\ref{eq59}), the latter remains valid as long as $P_k(\sigma,t)$ is sufficiently away from the boundary until the end of inflation. This situation, however, describes the case when $\sigma$ remains in slow-roll in the entire observable Universe and, as already stated, this is not the situation of interest to us. Instead, to treat the case when out-of-equilibrium patches appear in sparse regions of the Universe we must describe the evolution of $P_k(\sigma,t)$ as field interactions start to become dynamically important, namely when $P_k(\sigma,t)$ traverses the barrier at $\sigma_c$. The difficulty in this case is that the evolution can become significantly complicated. For example, if $P_k(\sigma,t)$ reaches the barrier for $t<t_k$, then ${\cal D}_k={\cal D}$, and the probability density obeys the usual Fokker-Planck equation in Eq.~(\ref{eq15}). As a result, the Gaussian solution in Eq.~(\ref{eq29}) becomes distorted in the neighborhood of the absorbing barrier in order to comply with the boundary condition Eq.~(\ref{eq26}) \cite{Sanchez:2012tk}. However, the behavior of $P_k(\sigma,t)$ is very different when it reaches the barrier for $t>t_k$. Since ${\cal D}_k=0$ in that case, the modified Fokker-Planck equation becomes a first order one in $\sigma$, thus admitting propagating solutions in just one direction. This implies that $P_k(\sigma,t)$ becomes ``absorbed'' by the barrier at $\sigma=\sigma_c$ without undergoing any distortion. It can be verified numerically that as the diffusion coefficient decreases, i.e. for ${\cal D}\ll H^3/4\pi^2$, the solution to the usual Fokker-Planck equation Eq.~(\ref{eq15}) with the boundary condition Eq.~(\ref{eq26}) approaches the solution with ${\cal D}_k=0$. In particular, we checked this agreement numerically down to values as small as ${\cal D}=10^{-2}H^3/4\pi^2$. We thus conclude that the solution in Eq.~(\ref{eq29}) holds exactly until the end of inflation in the physical region $\sigma\geq\sigma_c$ (thus affording us to use Eq.~(\ref{eq59})) provided $P_k(\sigma,t)$ reaches the barrier for $t>t_k$, namely after the scale $k^{-1}$ becomes superhorizon. Below we show how this consistency condition can be satisfied for the range of scales probed in the CMB if $\sigma_*$ is sufficiently larger than $\sigma_c$.

\subsection{Consistency condition for CMB scales} 
Once CMB scales are outside the horizon, the probability $P_k(\sigma,t)$ still has to evolve for a few tens of $e$-foldings before the end of inflation. During this time, the mean field $\bar\sigma$ decreases by a factor $\exp\left[-c_\sigma(N_*-N_{\rm CMB})/3\right]$, where $N_{\rm CMB}\simeq9$ is the number of $e$-foldings necessary for CMB scales to cross outside the horizon. In our range of interest $c_\sigma={\cal O}(10^{-1})$, such a factor can become larger than unity, thus implying a significant evolution of $P_k(\sigma,t)$. As a result, if field interactions are already important when CMB scales exit the horizon, namely $\bar\sigma\sim\sigma_c$, the expected outcome is that the most part of $P_k(\sigma,t)$ has been absorbed by the barrier before the end of inflation. In that case, $\sigma$ finds itself oscillating in the entire Universe at the end of inflation, thus becoming too small to have any observational consequence. The simplest manner to avoid this situation consists in setting $\sigma_*$ sufficiently large. Note that this requires certain amount of tuning, for too large a value of $\sigma_*$ might equally preclude the emergence of out-of-equilibrium patches at the end of inflation. For the time being, we analyze the consistency condition that affords us to use Eq.~(\ref{eq59}), deferring the discussion on parameter tuning to the Appendices~\ref{sec12} and \ref{sec11}. 

To establish the consistency condition validating the use of Eq.~(\ref{eq59}) we utilize the variable $\xi(k,t)$ introduced in Eq.~(\ref{eq3}). Then, most of the probability $P_k(\sigma,t)$ is far from the boundary at $\sigma_c$ as long as $\xi(k,t)>{\cal O}(1)$. For practical purposes, it  suffices to consider $\xi\geq2$, which corresponds to more than 99\% of the field distribution above the barrier. Using Eq.~(\ref{eq30}) and that $N\leq N_{\rm CMB}$ while CMB scales are crossing outside the horizon, we obtain the bounds $\bar\sigma(t)\geq\sigma_*\exp[-c_\sigma N_{\rm CMB}/3]$ and \mbox{$\Sigma_k^2(t)\leq\frac{3H^2}{8\pi^2c_\sigma}\left(1-\exp\left[-2c_\sigma N_{\rm CMB}/3\right]\right)$}. Substituting into the definition of $\xi$ we obtain the corresponding lower bound, which grows with the ratio $\sigma_*/\sigma_c$, as expected. As an example, for any $g\leq{\cal O}(1)$ and $c_\sigma={\cal O}(10^{-1})$ we find that it suffices to take $\sigma_*/\sigma_c\leq3$ to satisfy $\xi\geq3$ (more than 99\% of $P_k(\sigma,t)$ above the barrier) until CMB scales exit the horizon. Nevertheless, we notice that the ratio $\sigma_*/\sigma_c$ cannot be arbitrarily large, for $\rho_\sigma$ must remain subdominant during inflation. Using $\sigma_c\sim g^{-1}H_*$ to rewrite Eq.~(\ref{eq60}) as an upper bound for the ratio $\sigma_*/\sigma_c$ we obtain
\[
\frac{\sigma_*}{\sigma_c}<gc_\sigma^{-1/2}\frac{m_P}{H_*}\exp\left(c_\sigma N_{\rm sr}^p/3\right)\,.
\]
The strongest constraint is obtained when the right-hand side is at its minimum, for $c_\sigma=\frac{3}{2N_{\rm sr}^p}$. In that case we find
\[
\frac{\sigma_*}{\sigma_c}<g\sqrt{N_{\rm sr}^p}\,\frac{m_P}{H_*}\,,
\]
which leaves plenty room for the above choice $\sigma_*/\sigma_c\leq3$ for most values of $g$. In turn, values of $\sigma_*/\sigma_c$ within the same order of magnitude and for $g\geq10^{-2}$ can be easily obtained from a fast-roll stage with $0.2\lesssim\epsilon\lesssim0.3$ and $15\lesssim N_{\rm fr}\lesssim 20$.

\section{A first look at parameter tuning}\label{sec12}
Before computing $I$ in Eq.~(\ref{eq74}) we recall that in our setting inflation begins with $\sigma=0$ and that, following the discussion in Sec.~\ref{sec13}, we consider a sustained phase of fast-roll (characterized by a constant $\epsilon$ and lasting for $N_{\rm fr}$ $e$-foldings) during which the field variance $\Sigma^2$ undergoes an unstable growth due to the rapid evolution of the background, according to Eq.~(\ref{eq10}). After that, the fast-roll stage gives way to $N_{\rm sr}$ $e$-foldings of slow-roll inflation.

To clarify our discussion, in Fig.~\ref{fig17} we illustrate the probability density $G(\sigma)$ along with a number of possible situations (cases I to III) leading to the formation of patches at the end of inflation. In case I, $N_{\rm fr}$ and $\epsilon$ are such that the growth of $\Sigma$ during the fast-roll stage is insufficient for $G$ to encompass the range $\Delta\sigma$ within the expected region. The integral $I$ in this case becomes exponentially small. In case II, the growth undergone by $\Sigma$ is enough so that the central region of $G$ encompasses the range $\Delta\sigma$, with the latter still comparable to $\Sigma$. In this instance, $I$ can become of order 1, and hence the generation of patches becomes an expected outcome. Case III arises when $\Sigma$ grows much larger than $\Delta\sigma$. As a result, the formation of patches requires a considerable tuning of the initial value $\sigma_{\rm sr}$, since this must be confined to a very narrow interval.
\begin{figure}[htbp]
\centering\epsfig{file=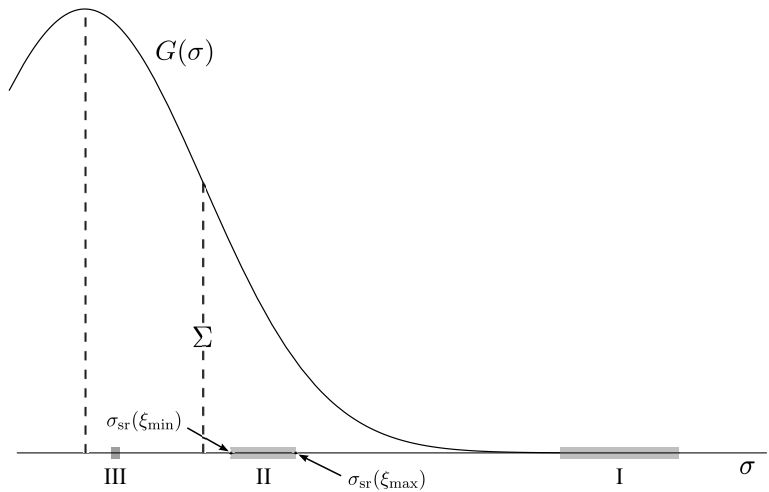,width=10cm}\caption{Field distribution $G(\sigma)$ at the onset of slow-roll inflation. Regions I, II and III represent different cases leading to the formation of out-of-equilibrium patches at the end of inflation.}\label{fig17}
\end{figure}

Here, an important point to stress is that $\Sigma$ depends on the inflationary model through $N_{\rm fr}$ and $\epsilon$. Therefore, the required model must be able to accommodate an epoch of fast-roll inflation ending well before cosmological scales exit the horizon and generating an initial condition for $\sigma$ compatible with the emergence of out-of-equilibrium patches. Further imposing the agreement of the subsequent slow-roll inflation with observations one can find the allowed range for $N_{\rm fr}$ and $\epsilon$, and from these, the allowed range of parameters of the inflationary model. This is precisely how our framework can be used as a tool to discriminate models of inflation: by demanding that $I$ in Eq.~(\ref{eq74}) is not too small. Although models satisfying this requirement can be presumably built, their construction and further exploration is beyond the scope of this paper and, consequently, the scenario here presented lacks the necessary input to evaluate the above integral. 

Despite this rather inconclusive statement, it is still possible to obtain valuable information to assess whether the emergence of out-of-equilibrium patches is a likely outcome. To see this, first we need to compute $\Delta\sigma$. According to Eq.~(\ref{eq3}), to constrain ${\cal F}(k)$ to a given range $[{\cal F}_{\rm min},{\cal F}_{\rm max}]$ we impose that at the end of inflation $\xi$ satisfies
\[\label{eq71}
\xi_{\rm min}\leq\xi(k,t_{\rm end})\leq\xi_{\rm max}\,,
\]
where $\xi_{\rm min}$ [$\xi_{\rm max}$] is the smallest [largest] value of $\xi$ compatible with the formation of patches with ${\cal F}(k)$ in its aforementioned interval. At the end of inflation, the mean field $\bar\sigma(t_{\rm end})$ corresponding to a given $\xi$ can be obtained using the definition of $\xi(k,t)$ introduced in Eq.~(\ref{eq3}). Using now $\bar\sigma(t_{\rm end})=\sigma_*e^{-c_\sigma N_*/3}$ with $\sigma_*=\sigma_{\rm sr}\exp(-c_\sigma N_{\rm sr}^p/3)$ we find the corresponding $\sigma_{\rm sr}$
\[\label{eq75}
\sigma_{\rm sr}(\xi(k,t_{\rm end}))=\left[\sigma_c+\sqrt2\,\Sigma_k(t_{\rm end})\xi(k,t_{\rm end})\right]\exp(c_\sigma N_{\rm sr}/3)\,,
\]
and hence the interval $\Delta\sigma$ corresponding to Eq.~(\ref{eq71})
\[\label{eq48}
\Delta\sigma=\sqrt2\,\Sigma_k(t_{\rm end})\Delta\xi\,\exp(c_\sigma N_{\rm sr}/3)\,,
\]
where $\Delta\xi\equiv\xi_{\rm max}-\xi_{\rm min}$.

Clearly, to avoid case I (see Fig.~\ref{fig17}) we must require that $\sigma_{\rm sr}(\xi_{\rm min})$ lies within the expected region, i.e. $\sigma_{\rm sr}(\xi_{\rm min})\lesssim\Sigma$. Therefore, owing to the exponential factor in Eq.~(\ref{eq75}), a large value of $c_\sigma N_{\rm sr}$ quickly tends to make the emergence of patches an unlikely event, unless this is conveniently counteracted by a similar growth of $\Sigma$ during the fast-roll stage. In this sense, and to minimize our demands from the fast-roll stage, it is desirable to stick to values of $N_{\rm sr}$ somewhat larger than $N_*$, but not much larger so as to make the emergence of patches unlikely, as in case I. In any case, we recall that the constraint in Eq.~(\ref{eq44}) must be satisfied.

To go beyond general conclusions and make quantitative statements about the  integral in Eq.~(\ref{eq74}) we need to make some assumptions regarding $\Sigma$. For example, a meaningful question to investigate is how likely is the appearance of patches, provided the appropriate conditions for that to happen are a priori assumed. Based on our previous discussion, by appropriate conditions here we mean that $\sigma_{\rm sr}(\xi_{\rm min})\sim\Sigma$. As illustrated in Fig.~\ref{fig17}, cases II and III satisfy this requirement, but only in case III the emergence of out-of-equilibrium patches becomes very unlikely. Therefore, we want to find out how model parameters determine which of the two cases arises. Taking $\Sigma=\sigma_{\rm sr}(\xi_{\rm min})$ for concreteness we can finally compute $I$, obtaining Eq.~(\ref{eq33}).

\section{Intersecting the last scattering surface}\label{app2}
To obtain an estimate of the probability $P_{\rm lss}(k)$ we assume that the \mbox{$k$-patches} that emerge at the end of inflation are randomly located in the observable Universe. Since the formation of the classical field is the result of the translation-invariant particle production mechanism undergone by $\sigma$, this is a reasonable assumption to make. Nevertheless, we notice that the existence of field correlations on scales up to ${\cal H}_*^{-1}$ can make the locations of out-of-equilibrium patches not completely random. Hence, on general grounds one can expect that the distribution of $k$-patches will feature certain degree of clustering or scale-dependence.

An appropriate measure of probability consists in identifying the fraction of volume (in the observable Universe) occupied by a given region with the probability that a random location in the observable Universe falls inside that region. Resorting to this interpretation, the probability that a sphere of radius $r$ falls entirely within the last scattering surface corresponds to the probability that their centers are separated by a distance smaller than $r_{\rm lss}-r$, namely
\[\label{eq29b}
P_{\rm in}(r)=\frac{\frac43\pi (r_{\rm lss}-r)^3}{(2r_{\rm lss})^3}=\frac{\pi}6\left(1-\frac{r}{r_{\rm lss}}\right)^3\,,
\]
which can be used in the interval  $0\leq r\leq r_{\rm lss}$. For $r>r_{\rm lss}$, this probability is obviously zero. Similarly, a sphere of radius $r$ falls entirely outside the last scattering surface when their centers are at a distance larger than $r_{\rm lss}+r$, and hence the associated probability is
\[\label{eq30b}
P_{\rm out}=1-\frac{\pi}6\left(1+\frac{r}{r_{\rm lss}}\right)^3\,,
\]
which can be used for $r\leq0.24r_{\rm lss}$ to keep a non-negative probability. Using Eqs.~(\ref{eq29b}) and (\ref{eq30b}), the probability that a sphere of radius $r$ intersects the last scattering surface is
\[\label{eq31b}
P_{\rm lss}(r)=1-(P_{\rm in}+P_{\rm out})=\frac13\,\pi x\left(3+x^2\right)\,\,,\,\,x\equiv\frac{r}{r_{\rm lss}}\,.
\]
To express $x$ in terms of the ratio $k/{\cal H}_*$, we denote the comoving scale currently entering the horizon by \mbox{$k_{\rm hor}={\cal H}_0={\cal H}_*$}. Using now $k_{\rm lss}=r_{\rm lss}^{-1}\simeq{\cal H}_0/2$ we find
\[
x=\frac{k_{\rm lss}}{k}=\frac{{\cal H}_*}{k}\frac{k_{\rm lss}}{{\cal H}_0}\simeq\frac{{\cal H}_*}{2k}\,.
\]
Substituting the above into Eq.~(\ref{eq31b}) we obtain Eq.~(\ref{eq83}).

\section{Evolution beyond the absorbing barrier}\label{sec16}
A simple alternative towards a physically motivated density $P_k^{\rm (ph)}(\sigma,t)$ relies on its construction as superposition of propagating $\delta$-like impulses sourced by $P_k(\sigma,t)$. At a given time $t$, the $\delta$-like impulses used to build $P_k^{\rm (ph)}(\sigma,t)$ are continuously generated at $\sigma=\sigma_c$ at all times $\tau\leq t$ with the initial amplitude $P_k(\sigma_c,\tau)$. Multiplying the latter times $\theta(\sigma-\sigma_c)$ to account for the conversion of the initial density $P_k(\sigma,t)$ into propagating impulses, the resulting extended probability density is
\[\label{eq76}
P_k^{\rm (ext)}(\sigma,t)=\theta(\sigma-\sigma_c)P_k(\sigma,t)+P_k^{\rm (ph)}(\sigma,t)\,.
\]
    
In this setting, the transition to the oscillatory phase is dictated by the dynamics of the $\delta$-like impulses. Regarding their propagation, we choose to keep our approach in the simplest and take the location of the $\delta$-like impulses to evolve exactly as the mean field $\bar\sigma$ during the slow-roll [c.f. Eq.~(\ref{eq30})]. At any time $t$, the location of the pulse generated at $\tau\leq t$ is then
\[
\sigma_p(t;\tau)=\sigma_c\exp\left[-c_\sigma H_*(t-\tau)/3\right]\,.
\]
Clearly, this is only approximate since the growth of the effective mass $m_\sigma$ entails a field evolution faster than during slow-roll. Nevertheless, this approximation facilitates a simple analytical solution for $P_k^{\rm (ph)}(\sigma,t)$ and, moreover, is sufficient for illustration purposes. Apart from the location of the impulses, we need to model the steady depopulation of the slow-roll phase as the field $\sigma$ enters its oscillatory stage. To do so, we introduce a timescale $\tau_t$ to parametrize the duration of the transition from the slow-roll to the oscillatory stage. We treat $\tau_t$ as a free parameter subject to the condition $H_*\tau_t\geq{\cal O}(1)$. Based on the stochastic nature of the particle production process responsible for the transition, the simplest alternative to account for this depopulation is by damping $P_k^{\rm (ph)}(\sigma,t)$ in the timescale $\tau_t$. Since $P_k^{\rm (ph)}(\sigma,t)$ is built as a superposition of impulses, this damping is naturally accounted for by imposing that the amplitude of the $\delta$-like impulses decreases exponentially in this timescale. Therefore, at any time $t$, the $\delta$-like impulse generated at $\tau\leq t$ is given by
\[\label{eq35}
p(\sigma,t;\tau)=\theta(t-\tau)\delta(\sigma-\sigma_p)P_k(\sigma_c,\tau)
\exp\left[-(t-\tau)/\tau_t\right]\,.
\]
Integrating now the collection of impulses with $\tau\leq t$ we obtain
\[\label{eq70}
P_k^{\rm (ph)}(\sigma,t)=\int_0^tp(\sigma,t;\tau)\,d\tau=\theta(\sigma_c-\sigma)P_k(\sigma_c,\tau_p)
\left(\frac{\sigma}{\sigma_c}\right)^{\frac3{c_\sigma H_*\tau_t}}\,,
\]
where
\[
\tau_p(\sigma,t)=t+\frac{3}{H_*c_\sigma}\ln\frac{\sigma}{\sigma_c}\,.
\]
Since $P_k^{\rm (ph)}(\sigma,t)$ describes the depopulation of the slow-roll phase, it entails a probability loss, which is then compensated by the corresponding buildup of probability in the oscillatory stage around $\sigma=0$. Although this buildup is unaccounted for in our approach, such an omission is relatively unimportant since $\sigma$ becomes exponentially suppressed (with respect to $\sigma_c$) during the oscillatory phase. 

\begin{figure}[htbp]
\centering \epsfig{file=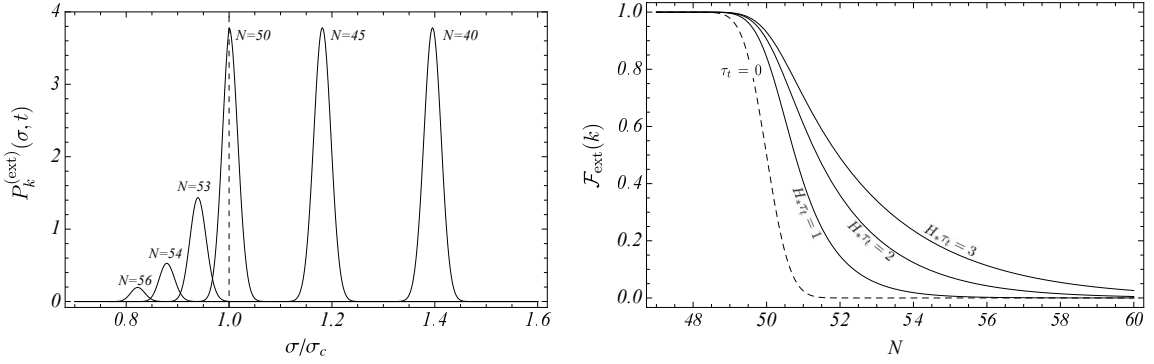,width=15.0cm}\caption{Plot of the extended probability density $P_k^{\rm (ext)}(\sigma,t)$ for $k=e^{N_{\rm CMB}}{\cal H}_*$ and $N$ $e$-foldings after the largest cosmological scales exit the horizon (left-hand panel). Plot of the corresponding fraction ${\cal F}_{\rm ext}(k)$ for different values of $\tau_t$, as indicated (righthand panel).}\label{fig14}
\end{figure}
In Fig.~\ref{fig14} (left-hand panel) we plot the behavior of the extended density $P_k^{\rm (ext)}(\sigma,t)$ for $k=e^{N_{\rm CMB}}{\cal H}_*$, thus allowing the emergence of out-of-equilibrium patches on all CMB scales, and at different times, before and after field interactions become important. We take $c_\sigma=0.1$ and $\sigma_{\rm sr}/\sigma_c\simeq5.3$ so that the probability $P_k^{\rm (ext)}(\sigma,t)$ reaches the absorbing barrier around the $e$-folding $N=50$ after the largest cosmological scales exit the horizon. Also, we consider the transition timescale $\tau_t=2H_*^{-1}$. The different curves depicted correspond to snapshots of $P_k^{\rm (ext)}(\sigma,t)$ taken at different number of $e$-foldings $N$. Our plot shows the steady depopulation of the slow-roll phase once interactions become important. Or equivalently, the persistence of the probability $P_k^{\rm (ext)}(\sigma,t)$ below the absorbing barrier for times of order $\tau_t$. In the righthand panel we depict the corresponding extended fraction ${\cal F}_{\rm ext}(k)$ for different choices of the transition timescale, $H_*\tau_t=1,2,3$. For comparison, we include the predicted fraction ${\cal F}(k)$ for an absorbing barrier\footnote{Note that in the limit of a very fast transition to the oscillatory stage, i.e. for $\tau_t\to0$, the case of an absorbing barrier is trivially recovered since \mbox{$P_k^{\rm (ph)}\to0$}.} (dashed line). As anticipated, the absorbing barrier approximation clearly underestimates the abundance of out-of-equilibrium patches. Our plot evidences how the temporary survival of the extended probability $P_k^{\rm (ext)}$ below $\sigma_c$ delays the suppression of ${\cal F}_{\rm ext}(k)$, thus allowing the persistence of out-of-equilibrium patches in sparse regions of the Universe for longer times. In turn, this tends to make their emergence more natural. This important point is further discussed below.

\subsection{Parameter tuning revisited}\label{sec11}
To estimate the level of tuning it will be useful to reproduce the result in Eq.~(\ref{eq33}) in a different manner. To recompute $I$ we first obtain the range $\Delta\sigma$ in Eq.~(\ref{eq48}) using the time lapse during which $\xi_{\rm min}\leq\xi(k,t)\leq\xi_{\rm max}$. Defining $t_{\rm max}$ and $t_{\rm min}$ as the time when $\xi=\xi_{\rm max}$ and $\xi=\xi_{\rm min}$, respectively, the length of this lapse is $\Delta t\equiv t_{\rm max}-t_{\rm min}$. Also, we must have $t_{\rm min}\leq t_{\rm end}\leq t_{\rm max}$ so that out-of-equilibrium patches emerge with a fraction within the limits determined by $\xi_{\rm min}$ and $\xi_{\rm max}$. Moreover, we may assume that during this lapse $P_k(\sigma,t)$ moves at an approximately constant speed, given by the mean velocity $\dot{\bar\sigma}$ evaluated at $\bar\sigma=\sigma_c$. This approximation holds whenever the width of $P_k(\sigma,t)$ satisfies $\Sigma_k(t)<\sigma_c$ for $t$ in the interval $[t_{\rm min},t_{\rm max}]$. But since $\Sigma_k$ decreases for $t>t_k$ (see Eq.~(\ref{eq30})), the condition $\Sigma_k(t)<\sigma_c$ for $t_{\rm min}\leq t_{\rm end}\leq t_{\rm max}$ implies\footnote{In fact, it can be shown that $\Sigma_k(t)$ remains nearly constant if $\Sigma_k(t)<\sigma_c$ for $t_{\rm min}\leq t\leq t_{\rm max}$. Since $P_k(\sigma,t)$ is crossing the barrier during this lapse, we can write $\sigma=\bar\sigma+\Delta\sigma$, with $\bar\sigma\simeq\sigma_c$ and $|\Delta\sigma|\sim\Sigma_k$. Therefore, assuming that $\Sigma_k(t)<\sigma_c$ implies that $P_k(\sigma,t)$ needs to move only a small amount in field space, namely $\bar\sigma(t_{\rm max})\simeq\bar\sigma(t_{\rm min})$, so that $\xi$ crosses the interval $[\xi_{\rm min},\xi_{\rm max}]$. Using now that both $\bar\sigma$ and $\Sigma_k$ decrease proportionally to $\exp(-c_\sigma N/3)$ we have
\[
\frac{\Sigma_k(t_{\rm min})}{\Sigma_k(t_{\rm max})}=\frac{\bar\sigma(t_{\rm min})}{\bar\sigma(t_{\rm max})}\simeq1\,,
\]
and hence $\Sigma_k(t)$ remains nearly constant.} $\Sigma_k(t_{\rm end})<\sigma_c$. Using the latter, we can approximate the length of the time lapse by
\[\label{eq73}
\Delta t\simeq\frac{\sqrt2\,\Sigma_k(t_{\rm end})\,\Delta\xi}{|\dot{\bar\sigma}(\bar\sigma=\sigma_c)|}\simeq\frac{\sqrt2\,\Sigma_k(t_{\rm end})\,\Delta\xi}{\sigma_c(c_\sigma H_*/3)}\,.
\]
From Eq.~(\ref{eq72}) we have
\[\label{eq49}
\sigma_{\rm sr}(\xi_{\rm max})=\sigma_{\rm sr}(\xi_{\rm min})\,\exp(c_\sigma H_*\,\Delta t/3)\,.
\]
Using $\Delta t$ in Eq.~(\ref{eq73}) and the a priori condition $\Sigma=\sigma_{\rm sr}(\xi_{\rm min})$ we find
\[\label{eq39}
\frac{\sigma_{\rm sr}(\xi_{\rm max})}{\Sigma}=\exp\left(\sqrt2\,\Sigma_k(t_{\rm end})\,\Delta\xi/\sigma_c\right)\,.
\]
Given that $\Sigma_k(t_{\rm end})/\sigma_c<1$ for natural values of $g$ and $c_\sigma$, we can Taylor expand the exponential, obtaining the result in Eq.~(\ref{eq46}) to first order in $c_\sigma$. 

The usefulness of this computation is twofold. On the one hand, it serves to illustrate the essential fact that $\Delta t$ determines the range of $\sigma_{\rm sr}$ compatible with the emergence of out-of-equilibrium patches with the designated abundances. On the other hand, since ${\cal F}_{\rm ext}(k)$ cannot be solved analytically, the above computation furnishes us with a method to estimate $I$ using $\Delta t$, which is directly related to transition timescale $\tau_t$ in our phenomenological model.

Now, to estimate $I$ we need to recompute $\Delta t$ taking into account the influence of the transition timescale $\tau_t$. Since the amplitude of the $\delta$-like impulses in Eq.~(\ref{eq35}) decreases by a factor of $e$ in a time $\tau_t$, its contribution to ${\cal F}_{\rm ext}(k)$ decreases by about one order of magnitude in a time $\sim2\tau_t$. Owing to this, the length of lapse during which $\xi$ traverses the interval $[\xi_{\rm min},\xi_{\rm max}]$ is enlarged, at least, by an amount of order $2\tau_t$. This can be appreciated too by inspecting the numerical results depicted in Fig.~\ref{fig14}. Therefore, by replacing
\[\label{eq85}
\Delta t\to\widetilde{\Delta t} =\Delta t+2\tau_t
\]
in Eq.~(\ref{eq49}) we recompute $\sigma_{\rm sr}(\xi_{\rm max})/\Sigma$, obtaining Eq.~(\ref{eq41}). An important aspect of this result is that when $c_\sigma$ is large enough so that $\widetilde{\Delta t}\simeq 2\tau_t$, the integral $I$ grows with $c_\sigma$. This happens because the exponential in Eq.~(\ref{eq41}) is dominated by the second term, thus implying the growth of $I$ with $c_\sigma$ (see Eq.~(\ref{eq33})). 

\bibliographystyle{h-physrev}
\bibliography{references}

\end{document}